\documentclass[12pt]{iopart}

\usepackage{iopams}

\usepackage{graphicx}
\usepackage{subfigure}%
\usepackage{dcolumn}
\usepackage{bm}
\usepackage{hyperref}
\usepackage{float}
\usepackage{threeparttable}%
\usepackage{amssymb}
\usepackage{amsgen}
\usepackage{amsfonts}
\usepackage{amsbsy}

\hypersetup{
     colorlinks   = true,
     citecolor    = blue
}

\hypersetup{colorlinks=true, citecolor=blue, linkcolor = blue, urlcolor = blue}

\begin{document}

\title[]{Magnetic Reconnection Process in Partially Ionized Fluids: Insights from the Solar Chromosphere}

\author{Adrian R. Montañez-Lobo$^{\dag}$ and F. D. Lora-Clavijo$^{\dag}$$^*$}
\address{$^{\dag}$ Grupo de Investigación en Relatividad y Gravitación, Escuela de Física, Universidad Industrial de Santander, A. A. 678, Bucaramanga 680002, Colombia}
\ead{fadulora@uis.edu.co}

\vspace{10pt}
\begin{indented}
\item[]July 2026
\end{indented}

\begin{abstract}
Magnetic reconnection converts stored magnetic energy into kinetic energy, heat, and radiation. While extensively studied in fully ionized plasmas, observations show that ionization and recombination also play a role by modifying the local plasma resistivity and enabling additional heating channels. This study examines magnetic reconnection in the partially ionized solar chromosphere, analyzing its morphology and energy releases with attention to elastic collisions, ionization, and recombination. The MAGNUS code, originally built for ohmic resistivity and heat transfer, was modified to handle elastic and inelastic collisions. The simulations are 2.5D resistive MHD with two-fluid effects (charged + neutrals), adapted to handle interactions via these collision terms. A mixed explicit-implicit scheme was implemented to manage the stiffness of these terms. Simulations were carried out for three different magnetic field strengths: $100$G, $110$G, and $120$G. These values correspond to the low to mid chromosphere in quiet Sun conditions. We found that reconnection heats the plasma components by $18\%$ to $110\%$. Although plasma beta rises only slightly, the energy release grows far more, indicating that charged–neutral collisions, not just beta or available magnetic energy, drive the enhanced reconnection. Furthermore, ionization and recombination become most significant in regions of peak temperature, particularly where particles are accelerated. Maximum reconnection rates from temporal analysis are $0.226$, $0.253$, and $0.279$, respectively. Finally, in terms of energy, our results show that in a chromospheric volume of $0.4 \times 0.01 \times 0.4$ Mm$^3$, the energy released ranges from $10^{22}$ to $10^{23}$.
\end{abstract}

%
\vspace{2pc}

\maketitle
%
%

\section{Introduction}

Partially ionized plasma is ubiquitous, found in molecular clouds \cite{mouschovias2011hydromagnetic}, planetary ionospheres \cite{ballester2018partially}, and protoplanetary disks \cite{kataoka2017ionization}. On Earth, it exists in the ionosphere and thermosphere, and in the Sun's chromosphere \cite{leake2014ionized}. In these regions, the plasma can be modeled as a fluid of neutral and charged particles \cite{judge2006observations}. Despite its known presence, the role of partially ionized plasma in major solar events has been overlooked. In this study, particular attention is given to solar flares, explosive events characterized by the release of large amounts of energy and the acceleration of solar material. These phenomena have been extensively documented, with magnetic reconnection identified as the primary mechanism driving energy release. Traditionally, solar flares have been primarily studied from a coronal perspective, as this region is assumed to contain the energy required to accelerate the plasma \cite{shibata2011solar}. Yet, evidence from ionization and recombination observations including hydrogen Balmer line emissions (e.g., $H_\alpha$, $H_\beta$) and continuum enhancements captured by instruments such as IRIS (Interface Region Imaging Spectrograph) and ground-based observatories \cite{2014SoPh..289.2733D,mulay2021evidence} suggests the role of partially ionized plasma in modifying the local heating rate, accelerating both ions and neutrals, and altering the thickness and dynamics of the reconnection current layer (e.g., \cite{Zweibel2011Magnetic,Ni2020Magnetic}), thereby affecting how magnetic energy is partitioned between charged and neutral species during flaring and sub-flaring events in the chromosphere. This has motivated particular interest in studying magnetic reconnection in the solar chromosphere, where partially ionized plasma is known to exist \cite{carlsson2019new}.

As previously mentioned, magnetic reconnection is the primary mechanism for solar flares, releasing stored magnetic energy through a restructuring of field lines \cite{pontin2022magnetic}. While extensively studied in fully ionized plasmas, reconnection in partially ionized plasmas has garnered significant interest due to its role in various astrophysical environments, including the solar chromosphere, planetary magnetospheres (e.g., Earth, Mars, Jupiter), and the interstellar medium \cite{Ni2020Magnetic, blanc2005solar, clark2023comprehensive}. Recent studies have modeled the chromosphere as separate charged and neutral fluids, assessing both whether reconnection can occur on solar flare timescales \cite{jara2019kinetic} and how collisions between species affect the process \cite{wargnier2023multifluid}.

Models have been proposed to explain the emission of ionization and recombination lines as Balmer lines, such as the chromospheric evaporation model. The latter consists of the upward expansion of the chromosphere after it is heated to coronal temperatures by either high‑energy particles or downward‑propagating conduction fronts from the coronal energy‑release site \cite{shibata2011solar}. The evaporated plasma fills the flare loops with hot, dense material, increasing temperature and coronal density, and potentially influencing the reconnection rate or the development of turbulence. Both then decrease as the plasma cools and drains back to the chromosphere, explaining observed spectra and Doppler shifts \cite{hirayama1974theoretical,wang2024spectral}. However, this model is challenged by persistent unexplained redshifts and flow velocities inconsistent with the associated energy in IRIS observations \cite{sadykov2015properties}, suggesting that, to achieve a more precise description of the magnetic reconnection phenomenon, it is appropriate to approach the problem through its description within the framework of partially ionized fluids. 
For this reason, particular attention is given to magnetic reconnection in the solar chromosphere, an environment composed of partially ionized plasma where ionization and recombination modify the local plasma resistivity, enable additional heating channels through collisional frictional interactions, and alter the reconnection rate, behaviors that are not captured by a single‑fluid model. The chromosphere is of additional interest as it represents a transition region. A manifestation of this behavior is observed in the plasma parameter plasma beta (which denotes the ratio between magnetic and kinetic pressures), 
which varies abruptly in this region, going from values of $10^2$ down to values of $10^{-4}$ \cite{gary2001plasma}, and in our work, presenting magnitudes from $10^0$ to $10^{-2}$, indicating that the influence of kinetic pressure becomes significant for the reconnection process.

Using a two-fluid model, this work quantifies how elastic and inelastic collisions influence magnetic reconnection in the solar chromosphere. For this purpose, a series of simulations are carried out in this work to investigate magnetic reconnection in the solar chromosphere, taking into account the stratification of this region. In the quiet Sun, typical magnetic field strengths range from 10 to 100~G, while in active regions they reach 150--1000~G \cite{pozuelo2023estimating,wang2026magnetic}. In this study, we work with magnetic field strengths of 100~G, 110~G, and 120~G. This relatively narrow range was chosen for a systematic investigation of the role of elastic and inelastic collisions on magnetic reconnection in a partially ionized plasma typical of the low to mid chromosphere under quiet Sun conditions. A broader parameter survey, including active region field strengths, is planned for future work. An extension of MAGNUS \cite{navarro2017magnus} is employed, a code adapted to solve the two-fluid equations. The objective is to quantify the energy released during the evolution of the phenomenon and to examine its morphological development, thereby assessing the behavior of reconnection in this region.

This article is structured as follows: Section \ref{sec:MHD} presents the two-fluid equations, including the models for elastic and inelastic collisions, and outlines the initialization of the chromospheric profiles. Section \ref{sec:RoC} investigates the collision frequencies and their atmospheric profiles. Section \ref{sec:NM} describes the numerical method, used to evolve the system, capturing the collision dynamics. Section \ref{sec:MRSC} analyzes the magnetic reconnection, including its morphology, rate, and associated energy conversion, alongside the evolution of the collision terms. Finally, Section \ref{sec:DaC} discusses the broader implications of our findings.

\section{Governing Equations for Partially Ionized Fluids}
\label{sec:MHD}

We model magnetic reconnection in a partially ionized plasma using a two-fluid framework. For this, we will use a hydrogen plasma, composed of two fluids. The first fluid is composed of the charged species (ions and electrons), while the second consists of the neutral species. This separation allows us to explicitly resolve their distinct dynamics and collisions. The system is evolved according to the two-fluid equations given in
\cite{ballester2018partially} and \cite{popescu2019two}:
\begin{eqnarray}
    &\partial_t \rho_n + \nabla \cdot (\rho_n \mathbf{u}_n) = S_n, \label{eq2.1} \\
    &\partial_t \rho_i + \nabla \cdot (\rho_i \mathbf{u}_i) = S_i, \label{eq2.2} \\
    &\partial_t (\rho_n \mathbf{u}_n) + \nabla \cdot (\rho_n \mathbf{u}_n \otimes \mathbf{u}_n + P_n \mathbf{I}) = \rho_n \mathbf{g} + \mathbf{R}_n, \label{eq2.3} \\
    &\partial_t (\rho_i \mathbf{u}_i) + \nabla \cdot \left[ \rho_i \mathbf{u}_i \otimes \mathbf{u}_i - \frac{\mathbf{B} \otimes \mathbf{B}}{\mu_0} + \left( P_i + \frac{\mathbf{B}^2}{2\mu_0} \right) \mathbf{I} \right] = \rho_i \mathbf{g} + \mathbf{R}_i, \label{eq2.4} \\
    &\partial_t E_n + \nabla \cdot \left[ (E_n + P_n) \mathbf{u}_n \right] = \rho_n \mathbf{g} \cdot \mathbf{u}_n + H_n, \label{eq2.5} \\
    &\partial_t E_i + \nabla \cdot \left[ \left( E_i + P_i + \frac{\mathbf{B}^2}{2\mu_0} \right) \mathbf{u}_i - \frac{\mathbf{B}}{\mu_0} (\mathbf{u}_i \cdot \mathbf{B}) \right] = - \nabla \cdot \left( \frac{\eta}{\mu_0} \mathbf{J} \times \mathbf{B} \right) \nonumber \\
    &\quad + \rho_i \mathbf{g} \cdot \mathbf{u}_i + H_i, \label{eq2.6} \\
    &\nabla \cdot \mathbf{B} = 0, \label{eq2.7} \\
    &\partial_t \mathbf{B} + \nabla \cdot (\mathbf{u}_i \otimes \mathbf{B} - \mathbf{B} \otimes \mathbf{u}_i) = - \nabla \times (\eta \mathbf{J}), \label{eq2.8} \\
    &\nabla \times \mathbf{B} = \mu_0 \mathbf{J}, \label{eq2.9}
\end{eqnarray}
where the subscripts $i$ and $n$ denote the charged and neutral species, respectively. The complete system comprises the mass conservation equations~\ref{eq2.1}--\ref{eq2.2}, momentum conservation~\ref{eq2.3}--\ref{eq2.4}, energy conservation~\ref{eq2.5}--\ref{eq2.6}, the magnetic solenoidal condition~\ref{eq2.7}, Faraday's induction law~\ref{eq2.8}, and Ampère's law in the low-speed approximation~\ref{eq2.9}. Here, $\rho_{i,n}$ are the mass densities, $\mathbf{u}_{i,n}$ the velocities, and $P_{i,n}$ the pressures. The identity tensor is $\mathbf{I}$, $\mathbf{g} = (0, -274, 0) \, \mathrm{m\,s^{-2}}$ is the solar gravitational acceleration, $\mathbf{B}$ the magnetic field, $\mathbf{J}$ the current density, $\eta$ the resistivity, and $\mu_0$ the permeability of free space. The total energy densities for the charged and neutral fluids are defined as:
\begin{eqnarray}
    E_i &=& \frac{1}{2} \rho_i u_i^2 + \frac{B^2}{2 \mu_0} + \rho_i e_i, \label{eq:energy_i} \\
    E_n &=& \frac{1}{2} \rho_n u_n^2 + \rho_n e_n, \label{eq:energy_n}
\end{eqnarray}
and the pressures for both fluids are given by the equations of state:
\begin{equation}
    P_{i,n} = (\gamma - 1) \rho_{i,n} e_{i,n} = \frac{k_B}{m_{i,n}} \rho_{i,n} T_{i,n}, \label{eq:pressure}
\end{equation}
where $T_{i,n}$ are the temperatures, $m_{i,n}$ the particle masses, $k_B$ the Boltzmann constant and $\gamma = 5/3$ the adiabatic index.

The terms  $S_{i,n}$, $\mathbf{R}_{i,n}$, and 
$H_{i,n}$ denotes the collisional source and sink rates for mass, momentum, and energy, respectively:
\begin{eqnarray}
    S_i &=& -S_n = \rho_{n}\Gamma^{\mathrm{ion}} - \rho_{i}\Gamma^{\mathrm{rec}}, \label{eq:Si} \\
    \mathbf{R}_i &=& -\mathbf{R}_n = \alpha (\mathbf{u}_n - \mathbf{u}_i) + \rho_{n} \mathbf{u}_n \Gamma^{\mathrm{ion}} - \rho_{i} \mathbf{u}_i \Gamma^{\mathrm{rec}}, \label{eq:Ri} \\
    H_i &=& -H_n = \frac{1}{2} \alpha \left(u_n^2 - u_i^2\right) + \frac{1}{2} \left(\rho_{n} u_n^2 \Gamma^{\mathrm{ion}} - \frac{1}{2} \rho_{i} u_i^2 \Gamma^{\mathrm{rec}}\right) \nonumber \\
    && + \frac{\alpha}{\gamma - 1} \frac{k_B}{\bar{m}} \left(T_n - T_i \right) + \frac{1}{\gamma - 1} \frac{k_B}{\bar{m}} \left( \rho_{n} T_n \Gamma^{\mathrm{ion}} - \rho_{i} T_i \Gamma^{\mathrm{rec}} \right), \label{eq:Hi}
\end{eqnarray}where $\bar{m}$ is the mean particle mass of the fluid, and $\Gamma^{\mathrm{ion}}$ and $ \Gamma^{\mathrm{rec}}$ are the ionization and recombination rates.

Equation~\ref{eq:Si} accounts for mass density exchange between ions and neutrals through ionization and recombination processes, thereby coupling the continuity equations of the two fluids. The momentum exchange between species is described by Equation~\ref{eq:Ri}, where the total collisional force density includes both elastic and inelastic contributions. The first term represents elastic ion–neutral collisions and arises from the relative velocity between the fluids. The other terms correspond to inelastic processes associated with ionization and recombination, where momentum is transferred as particles are converted from one species to the other.

The energy exchange is governed by Equation~\ref{eq:Hi}, which describes the collisional power density $H_{i,n}$. The first term represents the kinetic energy exchange associated with elastic collisions and corresponds to the work done by the collisional force appearing in $\mathbf{R}_{i,n}$. This can be made explicit by rewriting it as
\begin{equation}
\frac{1}{2}\alpha (u_n^2 - u_i^2)
= \alpha, \mathbf{u}_i \cdot (\mathbf{u}_n - \mathbf{u}_i)
+ \frac{1}{2}\alpha \left| \mathbf{u}_n - \mathbf{u}_i \right| ^2,
\end{equation}
which shows that the energy transfer consists of two distinct contributions: work performed by the elastic collisional force between the fluids, and irreversible frictional heating due to the relative drift between ions and neutrals. The second term in $H_{i,n}$ accounts for the kinetic energy exchanged through particle conversion during ionization and recombination. The third term describes thermal energy transfer driven by temperature differences between the two fluids, while the fourth term represents the thermal energy gained or lost through particle exchange. Together, these terms encapsulate the key collisional pathways through which momentum and energy are redistributed in partially ionized plasmas relevant to magnetic reconnection in the solar atmosphere.

The ionization and recombination rates, $\Gamma^{\mathrm{ion}}$ and $\Gamma^{\mathrm{rec}}$, are given by semi-empirical models \cite{1997ADNDT..65....1V,2003poai.book.....S}:
\begin{equation}
    \Gamma^{\mathrm{ion}} = 2.91 \times 10^{-14} \frac{n_e}{X + \phi_{\mathrm{ion}}/T_e^*} \left( \frac{\phi_{\mathrm{ion}}}{T_e^*} \right)^K e^{-\phi_{\mathrm{ion}}/T_e^*} \quad \mathrm{s^{-1}},
\end{equation}
\begin{equation}
    \Gamma^{\mathrm{rec}} = 2.6 \times 10^{-19} \frac{n_e}{\sqrt{T_e^*}} \quad \mathrm{s^{-1}},
\end{equation}
where $X = 0.232$ and $K = 0.39$ are fitting parameters, $T_e^*$ is the electron temperature in eV, $n_e$ is the electron number density, and $\phi_{\mathrm{ion}} = 13.6$ eV is the hydrogen ionization potential.

The collisional friction coefficient $\alpha$ is modeled as \cite{1965RvPP....1..205B,ballester2018partially}:
\begin{equation}
    \alpha = \frac{4}{3} \frac{\sigma \rho_i \rho_n}{m_i + m_n} \sqrt{\frac{8 k_B}{\pi} \left( \frac{T_i}{m_i} + \frac{T_n}{m_n} \right) },
    \label{eq:Alpha}
\end{equation}
where $\sigma$ is the effective collision cross-section. This expression derives from the reaction rate formalism of \cite{draine1986multicomponent}. For our system, initialized from rest and considering the high temperatures of the solar atmosphere, Equation~\ref{eq:Alpha} provides a suitable approximation. Although $\alpha$ can be treated as a free parameter to modulate elastic collision effects, we adopt a cross-section of $\sigma \sim 10^{-20}~\mathrm{m^2}$ based on established models \cite{1965RvPP....1..205B}.

\subsection{Modeling the Solar Background}

The initial hydrostatic equilibrium state for the charged particles is derived using the semi-empirical temperature model from \cite{2008ApJS..175..229A}. Combining the hydrostatic equilibrium equation with the ideal gas law,
\begin{equation}
    \nabla P_i = \rho_i \mathbf{g},
\end{equation}
\begin{equation}
    \rho_i = \frac{m_i}{k_B} \frac{P_i}{T_i},
\end{equation}
yields the pressure profile
\begin{equation}
    P_i = P_{i0} \exp\left(-\frac{m_i |\mathbf{g}|}{k_B} \int \frac{1}{T_i} \, dz\right),
\end{equation}
where the reference pressure is \( P_{i0} = 1.29 \, \mathrm{Pa} \), set in a atmospheric solar height of 5 Mm \cite{1981ApJS...45..635V,2021ApJ...911..119Z}. This solution follows from the momentum equation under the assumption of an initially force-free magnetic field (\(\mathbf{J} \times \mathbf{B} = 0\)).

\begin{figure}
    \centering
    \includegraphics[width=1.0\linewidth]{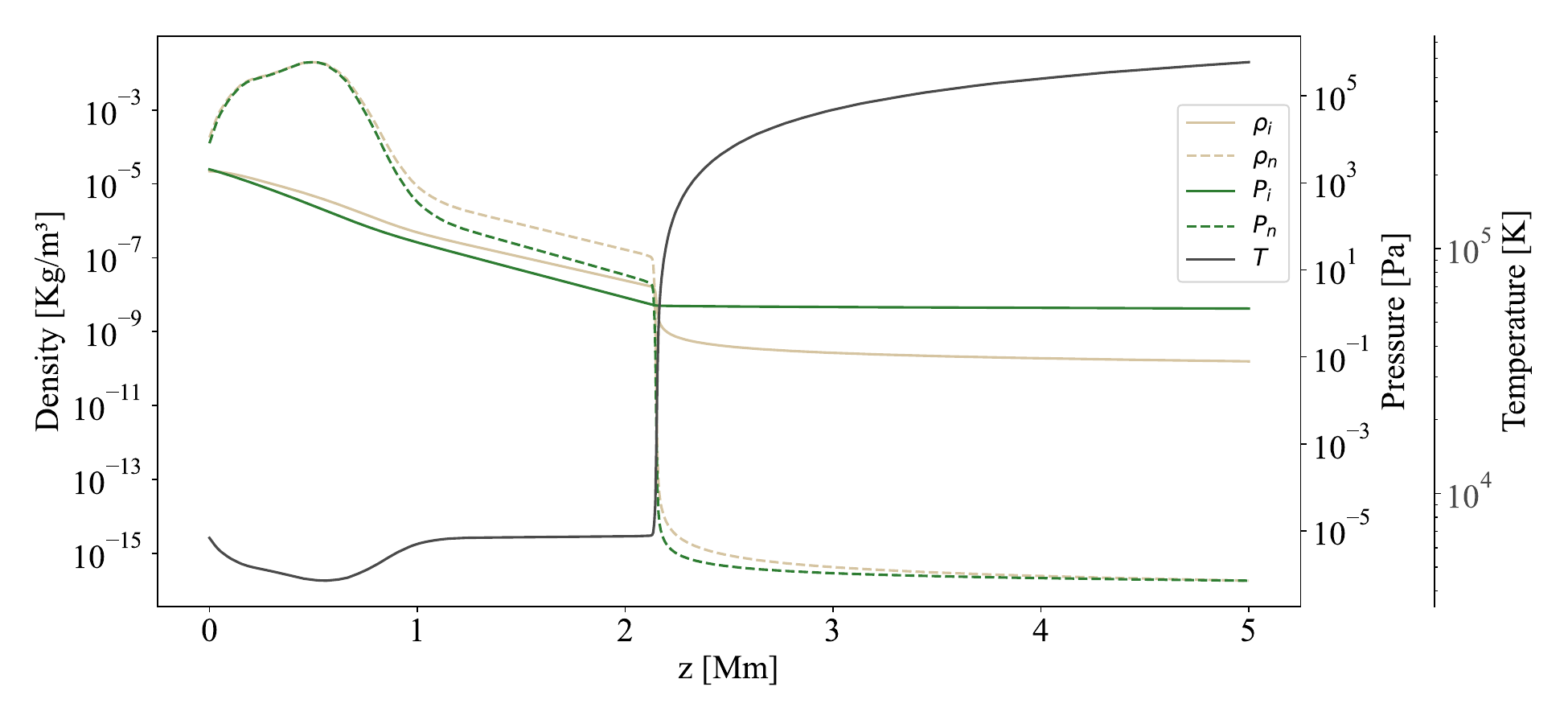}
    \caption{Representation of the equilibrium solar profile from the atmospheric base up to a height of h = 5 [Mm]. The temperature profile (black line) is shown with its corresponding values on the floating bar to the right, obtained from the observational models of Avrett. In addition, the interpolated profiles of the neutral species (dashed line) and the charged species (solid line) are displayed for the density profiles (beige) and the pressure profiles (green).}
    \label{fig:sp}
\end{figure}

For the neutral species, density and pressure profiles are determined by collisional processes. The equilibrium condition requires the collisional source terms for mass to be zero (\(S_i = S_n = 0\)), which enforces a local balance between ionization and recombination:
\begin{equation}
    \rho_n = \rho_i \frac{\Gamma^{\mathrm{rec}}}{\Gamma^{\mathrm{ion}}}. \label{eq4.1}
\end{equation}
The neutral pressure can then be obtained from the equation of state \ref{eq:pressure}. Similarly, the equilibrium state also demands that the collisional source terms for energy vanish ($H_i = H_n = 0$), which requires the neutral and ion temperatures to be equal: $T_n = T_i = T$. The resulting initial profiles for temperature, density, and pressure of both species are shown as functions of height in Figure \ref{fig:sp}.
It is worth mentioning that the neutral fluid is not in hydrostatic equilibrium, as the system is overdetermined: while neutrals satisfy the equation of state and the ionization-recombination balance \ref{eq4.1}, they do not satisfy the momentum balance \( \nabla P_n = \rho_n \mathbf{g} \). 

The initial magnetic field follows a force-free Harris current sheet \cite{1962NCim...23..115H}:
\begin{equation}
    B_z = -B_0 \tanh\!\left( \frac{x}{w_0} \right), \quad
    B_y = \sqrt{B_0^2 - B_z^2}, \quad
    B_x = 0,
\end{equation}
where \( B_0 \) corresponds to the mean chromospheric field strength from the VAL-C model \cite{1981ApJS...45..635V}, and \( w_0 \) is the current sheet transition width. Localized anomalous resistivity is specified as:
\begin{equation}
    \eta = \eta_0 \exp\!\left( -\frac{x^2 + (z - h_{\mathrm{rec}})^2}{w_\eta^2} \right),
\end{equation}
with \( h_{\mathrm{rec}} \) the reconnection height, \( w_\eta \) the resistivity-region width, $w_\eta = 0.066$\,Mm and \( \eta_0 = 6.28 \times 10^{4}~\Omega\!\cdot\!\mathrm{m} \). This amplitude produces Lundquist numbers \( S \sim 10^2-10^4 \), consistent with fast reconnection regimes \cite{2021PhPl...28c2901M}.
The anomalous resistivity locally seeds reconnection at a controlled location, reducing the numerical transient while preserving the physical dynamics. Its width ($0.066$\,Mm) is consistent with the expected thickness of a chromospheric current sheet in a partially ionized plasma \cite{zweibel2009magnetic}.

\begin{figure}
    \centering
    \includegraphics[width=0.8\linewidth]{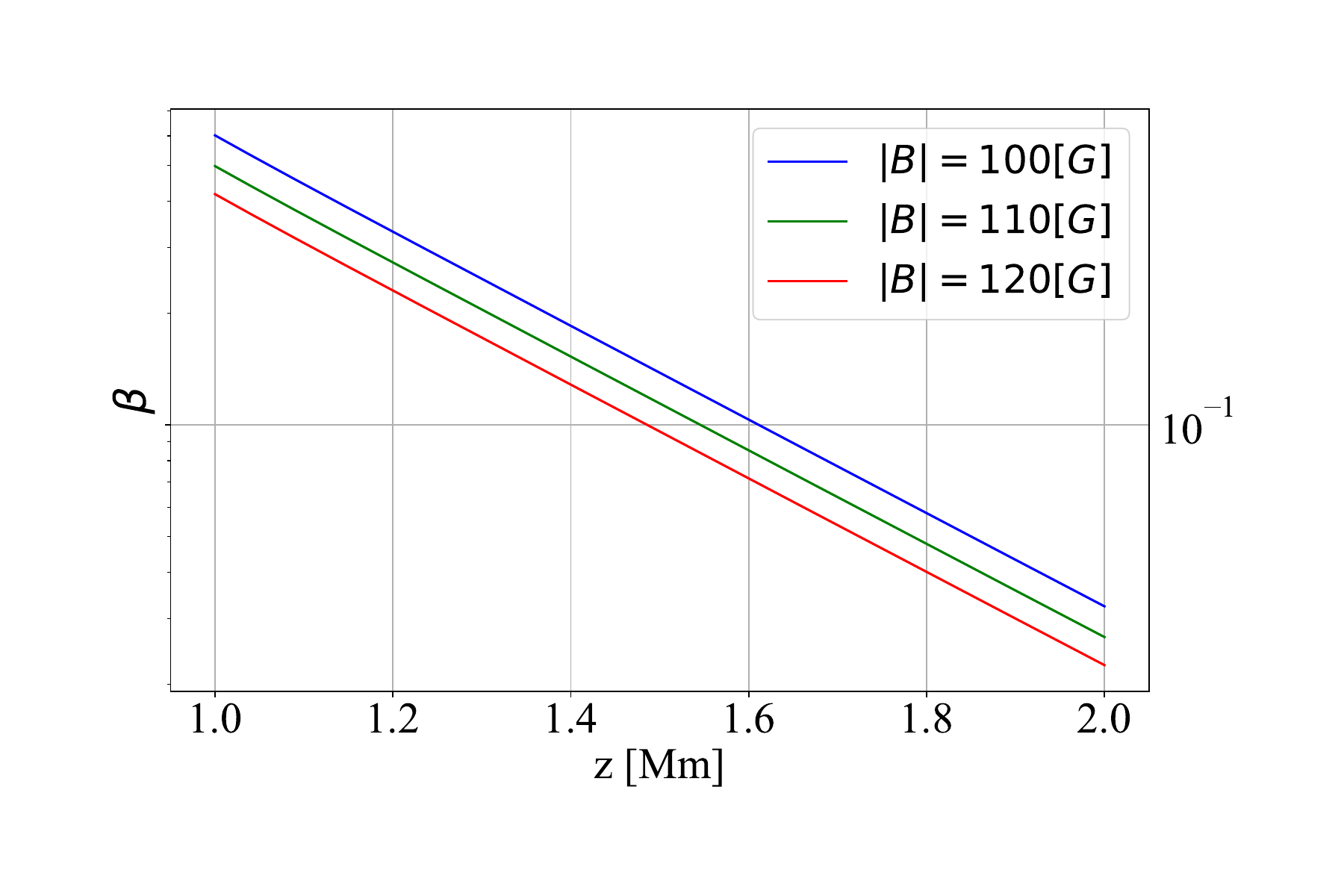}
    \caption{Plasma $\beta$ as a function of height $z$ (from 1 to 2 Mm) in the chromospheric region for three initial magnetic field strengths   $B_0$= 100 G (blue), 110 G (green), and 120 G (red). }
    \label{fig:bp}
\end{figure}

Finally, the behavior of the plasma beta parameter ($\beta$) in the equilibrium state must be defined. The plasma $\beta$ is calculated using the expression
\begin{equation}
    \beta = \frac{P_i}{B^2/2\mu_0},
\end{equation}
which defines the ratio between the kinetic and magnetic pressure in the ionized plasma. In Figure \ref{fig:bp} we can see the value of the plasma $\beta$ for the region between $z$ = 1 and 2 Mm. This corresponds to the chromospheric region, which is of interest for our simulations. Thus, the variation of the plasma $\beta$ is shown for the cases of initial magnetic field $B_0$ = 100 (blue), 110 (green), and 120~G (red). As can be seen, in all three cases, the parameter scales are within orders of $10^0$ to $10^{-2}$, within the observationally accepted values \cite{gary2001plasma}.

\section{The Role of Collision Terms in the Solar Atmosphere}
\label{sec:RoC}

Collision terms give rise to phenomena such as enhanced resistivity, prolonged reconnection times, and increased reconnection rates \cite{Zweibel2011Magnetic}. For this reason, it is essential to identify the regions of the solar atmosphere where ionization and recombination processes become dynamically significant. The present analysis is based on the solar atmospheric model shown in Figure~\ref{fig:sp}.

\begin{figure}
    \centering
    \includegraphics[width=0.8\linewidth]{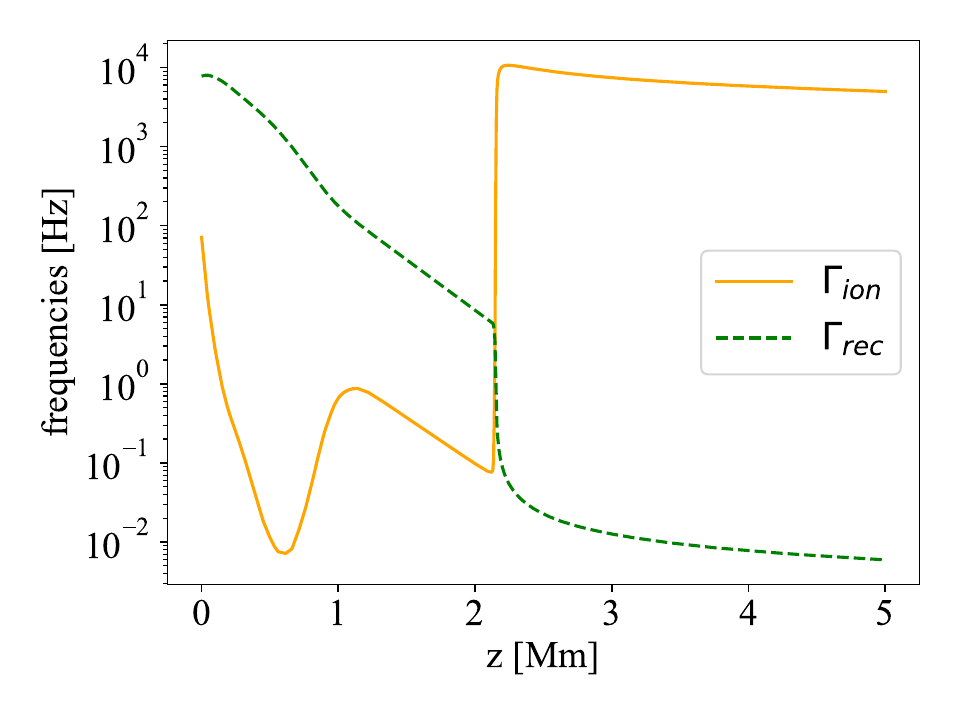}
    \caption{Variation of ionization (solid line) and recombination (dashed line) frequencies with height in the solar atmosphere.}
    \label{fig:AIF}
\end{figure}

Figure~\ref{fig:AIF} shows the ionization and recombination rates throughout the atmosphere. Below the transition region, recombination frequencies dominate over ionization frequencies, indicating that interactions between charged and neutral particles play a significant role in the plasma dynamics in this region. In contrast, above the transition region, the ionization frequency becomes dominant, exceeding the recombination frequency by several orders of magnitude. When this difference reaches approximately six orders of magnitude, the plasma may be regarded as fully ionized, marking a transition to a regime well described by single-fluid dynamics.

On the other hand, we examine the role of elastic collisions in coupling the dynamics of the charged and neutral fluids. To this end, we introduce the parameter $\alpha$, defined as
\begin{equation}
    \alpha = \rho_i \nu_{ni} = \rho_n \nu_{in},
    \label{eq:NiNn}
\end{equation}
where $\nu_{ni}$ and $\nu_{in}$ denote the elastic collision frequencies between neutral and charged particles. Here, $\nu_{ni}$ corresponds to collisions experienced by charged particles due to interactions with neutral species, while $\nu_{in}$ represents the reciprocal process. Expressing $\alpha$ in this form emphasizes that elastic collisions exert equal and opposite coupling forces on the two fluids, while their dynamical response is modulated by the inertia of each species through their respective mass densities.

\begin{figure}
    \centering
    \includegraphics[width=0.8\linewidth]{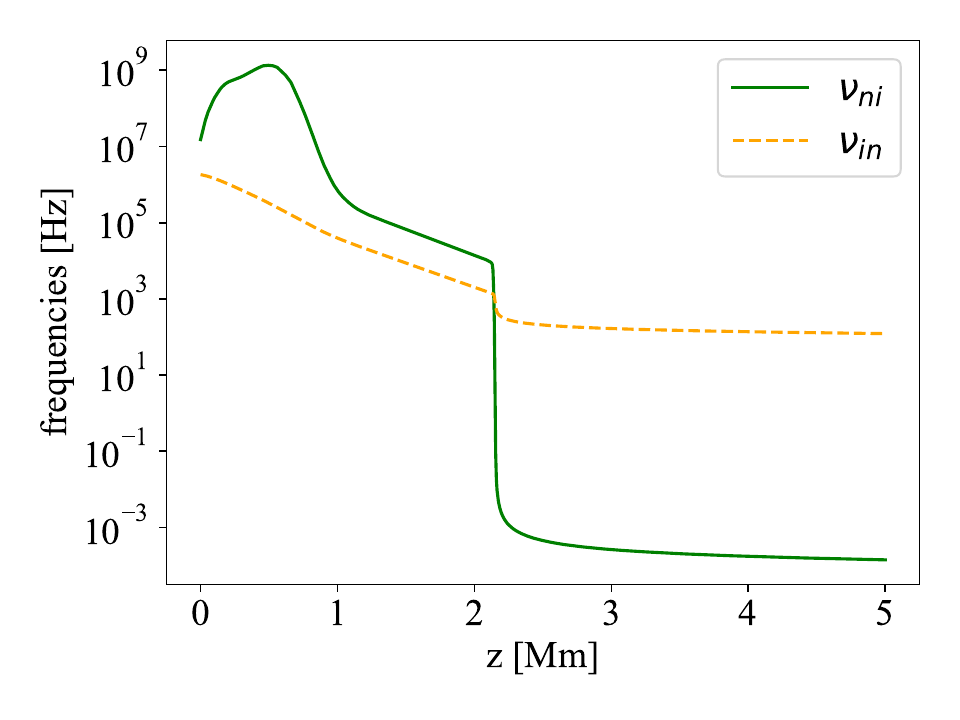}
    \caption{Effect of elastic collisions as a function of height. The graphic displays the collision frequency of neutral species with charged species ($\nu_{ni}$, green solid line) and the frequency of charged species with neutral species ($\nu_{in}$, orange dashed line).}
    \label{fig:AEF}
\end{figure}

Figure~\ref{fig:AEF} shows the height dependence of the elastic collision frequencies $\nu_{ni}$ and $\nu_{in}$. Throughout this work, including the calculation of these collision frequencies, we adopt a constant elastic collision cross section of $\sigma = 10^{-20}\,\mathrm{m}^2$. From Figure~\ref{fig:AEF}, three distinct regions can be identified. Below the chromosphere, the neutral--ion collision frequency $\nu_{ni}$ exceeds $\nu_{in}$ by approximately three orders of magnitude, reflecting the dominance of neutrals in this region. Above the transition region, the situation is reversed, with $\nu_{in}$ exceeding $\nu_{ni}$ by roughly six orders of magnitude, consistent with the predominance of the ionized component. In the chromosphere layer itself, the ratio between $\nu_{ni}$ and $\nu_{in}$ typically lies within one to two orders of magnitude, with $\nu_{ni} > \nu_{in}$ due to the higher neutral density. This asymmetry means that while ions are effectively coupled to neutrals via collisions, neutrals are not necessarily coupled to ions, nor are they coupled to the magnetic field. Consequently, the system does not behave as a single fluid; instead, the differential motion between ions and neutrals gives rise to frictional heating and modifies the reconnection dynamics in ways distinct from fully ionized.

In light of the considerations above, distinct dynamical regimes can be identified across the solar atmosphere. Below the chromosphere, the strong collisional coupling is dominated by the neutral component, and the dynamics of the charged species are largely controlled by neutral hydrodynamics. This region can therefore be regarded as primarily neutral in character. In contrast, above the transition region, the influence of neutrals on the charged species becomes negligible, and the plasma can be accurately described within the framework of fully ionized magnetohydrodynamics. 

Due to the predominantly neutral nature of the chromosphere, there are important physical consequences: neutrals frequently collide with ions, but ions are much less effective at perturbing the neutral population. As a result, the chromosphere maintains a regime where frictional heating is efficient, even while some degree of collisional coupling persists. This coexistence of neutral and charged effects, with the ionization fraction being the key parameter, defines the chromosphere as a genuine partially ionized plasma. As a result, both the hydrodynamic response of the neutral fluid and the magnetohydrodynamic behavior of the charged component play a central role in determining the system dynamics. Accordingly, we focus our analysis on the chromospheric layer, where collisional coupling between ions and neutrals is expected to exert a decisive influence on the onset and evolution of magnetic reconnection.

\section{Numerical Methods}
\label{sec:NM}

We use the MAGNUS code to numerically solve the equations for partially ionized fluids \cite{navarro2017magnus}. MAGNUS solves the resistive magnetohydrodynamics equations with finite volume methods and high-resolution shock-capturing schemes \cite{toro2013riemann}. It has previously been applied to solar physics problems, including plasma emergence into a coronal hole \cite{navarro2019numerical} and the formation of chromospheric jets with thermal conduction \cite{navarro2021thermal}. The code has also been used to study torsional Alfvén waves in a stratified atmosphere \cite{wandurraga2021torsional}. Recently, MAGNUS was extended to include the Hall effect and ambipolar diffusion, which are investigated as mechanisms for fast magnetic reconnection \cite{landinez2024systematic}.

For this study, we further adapt MAGNUS to solve two-fluid equations, treating charged and neutral species separately and coupling them through collisions. Because collision frequencies change abruptly across the sharply stratified solar atmosphere, these terms become stiff, especially in lower layers. To maintain stability under extreme temperature and pressure gradients, we implement an implicit-explicit (IMEX) scheme. This scheme is a numerical technique for solving differential equations that splits the equation into two components: a stiff \footnote{A system is stiff if the numerical method required for stability is forced to take much smaller time steps than the accuracy requirement alone would demand. In this case, the collision terms are considered stiff because their frequencies are sufficiently high compared to the plasma dynamics, especially in the chromosphere.} part treated with computationally stable implicit methods, and a non‑stiff part treated with highly efficient explicit methods. \cite{boscarino2024implicit}. This approach preserves stability for systems dominated by stiff terms. A Runge-Kutta (RK) subroutine handles the collision evolution, with an implicit matrix inversion of the Jacobian to manage stiffness, ensuring robust integration.

To integrate the governing equations in time, we employ an IMEX-RK scheme, which allows for a stable and efficient treatment of the stiff collisional source terms while retaining an explicit formulation for the remaining contributions. The system of evolution equations can be written compactly as
\begin{equation}
    \frac{d \mathbf{U}}{dt} = \mathbf{F}_{\mathrm{ns}}(\mathbf{U}) + \mathbf{F}_{\mathrm{s}}(\mathbf{U}),
\end{equation}
where $\mathbf{U}$ denotes the vector of conservative variables,
\begin{equation}
    \mathbf{U} = \left( \rho_i, \rho_i \mathbf{u}_i, E_i, \mathbf{B}, \rho_n, \rho_n \mathbf{u}_n, E_n \right).
\end{equation}
Here, $\mathbf{F}_{\mathrm{ns}}$ contains the non-stiff sources of the evolution, such as advective and magnetic contributions, while $\mathbf{F}_{\mathrm{s}}$ collects the stiff source terms arising primarily from collisional coupling between ions and neutrals.

Within an IMEX-RK, the solution is advanced from time level $n$ to $n+1$ according to
\begin{equation}
    \mathbf{U}^{n+1} = \mathbf{U}^{n} + \Delta t
    \left[
    \sum_{i=1}^s b_i \mathbf{F}_{\mathrm{ns}}(\mathbf{U}_i)
    + \sum_{i=1}^s \hat{b}_i \mathbf{F}_{\mathrm{s}}(\mathbf{U}_i)
    \right],
\end{equation}
where $s$ is the number of stages and $\mathbf{U}_i$ are the stage values, given by
\begin{equation}
    \mathbf{U}_{i} = \mathbf{U}^{n}
    + \Delta t \left[
    \sum_{j=1}^{i-1} A_{ij} \mathbf{F}_{\mathrm{ns}}(\mathbf{U}_j)
    + \sum_{j=1}^{i} \hat{A}_{ij} \mathbf{F}_{\mathrm{s}}(\mathbf{U}_j)
    \right].
\end{equation}
The coefficients $A_{ij}$, $\hat{A}_{ij}$, $b_i$, and $\hat{b}_i$ define the specific IMEX-RK scheme, with the non-stiff terms treated explicitly and the stiff terms handled implicitly.

For the stiff contribution, we adopt an explicit singly diagonally implicit Runge--Kutta (ESDIRK) scheme \cite{jørgensen2018}, in which the diagonal coefficients $\hat{A}_{ii}$ are non-zero. As a consequence, each stage requires the solution of an implicit system. The stage equation can be rearranged as
\begin{equation}
    \mathbf{U}_{i} =
    \mathbf{U}^{*}
    + \Delta t \sum_{j=1}^{i-1} \hat{A}_{ij} \mathbf{F}_{\mathrm{s}}(\mathbf{U}_j)
    + \Delta t \hat{A}_{ii} \mathbf{F}_{\mathrm{s}}(\mathbf{U}_i),
    \label{eq:stage_implicit}
\end{equation}
where the explicit predictor $\mathbf{U}^{*}$ gathers all contributions known from previous stages,
\begin{equation}
    \mathbf{U}^{*} =
    \mathbf{U}^{n}
    + \Delta t \sum_{j=1}^{i-1} A_{ij} \mathbf{F}_{\mathrm{ns}}(\mathbf{U}_j).
\end{equation}

To solve Eq.~\ref{eq:stage_implicit}, the stiff source term is linearized about the predictor state using a first-order Taylor expansion,
\begin{equation}
    \mathbf{F}_{\mathrm{s}}(\mathbf{U}_i)
    \approx
    \mathbf{F}_{\mathrm{s}}(\mathbf{U}^{*})
    + \mathbf{J} \cdot \left( \mathbf{U}_i - \mathbf{U}^{*} \right),
\end{equation}
where $\mathbf{J}$ is the Jacobian of the stiff source term evaluated at $\mathbf{U}^{*}$,
\begin{equation}
    \mathbf{J}
    =
    \left.
    \frac{\partial \mathbf{F}_{\mathrm{s}}(\mathbf{U})}{\partial \mathbf{U}}
    \right|_{\mathbf{U} = \mathbf{U}^{*}}.
\end{equation}
Substituting this approximation into Eq.~\ref{eq:stage_implicit} yields the final update for the stage solution,
\begin{equation}
    \mathbf{U}_{i} =
    \mathbf{U}^{*}
    + \left[ \mathbf{I} - \Delta t \hat{A}_{ii} \mathbf{J} \right]^{-1}
    \Delta t
    \left(
    \sum_{j=1}^{i-1} \hat{A}_{ij} \mathbf{F}_{\mathrm{s}}(\mathbf{U}_j)
    + \hat{A}_{ii} \mathbf{F}_{\mathrm{s}}(\mathbf{U}^{*})
    \right).
\end{equation}

The coefficients defining the explicit and implicit RK schemes are conveniently represented using Butcher tableaus \cite{butcher1996history}. For the non-stiff (explicit) part of the evolution, the tableau is
$$
\begin{array}{c|ccc}
0 & 0 & 0 & 0 \\
\omega & \omega & 0 & 0 \\
1 & \kappa & 1-\kappa & 0 \\
\hline
 & 0 & 1-\omega & \omega
\end{array}
$$
while the stiff (implicit) contribution is advanced using
$$
\begin{array}{c|ccc}
0 & 0 & 0 & 0 \\
\omega & 0 & \omega & 0 \\
1 & 0 & 1-\omega & \omega \\
\hline
 & 0 & 1-\omega & \omega
\end{array}.
$$
The parameters $\omega$ and $\kappa$ are free and control the balance between the explicit and implicit contributions, as well as the overall order and stability properties of the scheme. For the second-order IMEX formulation used in this work, the choice $\omega = (2-\sqrt{2})/2$ provides optimal linear stability for the stiff terms, while $\kappa = -2\sqrt{2}/3$ ensures second-order accuracy of the explicit Runge--Kutta scheme.

\section{Magnetic Reconnection at the Solar Chromosphere}
\label{sec:MRSC}

\begin{figure}
    \centering
\includegraphics[width=1.0\linewidth]{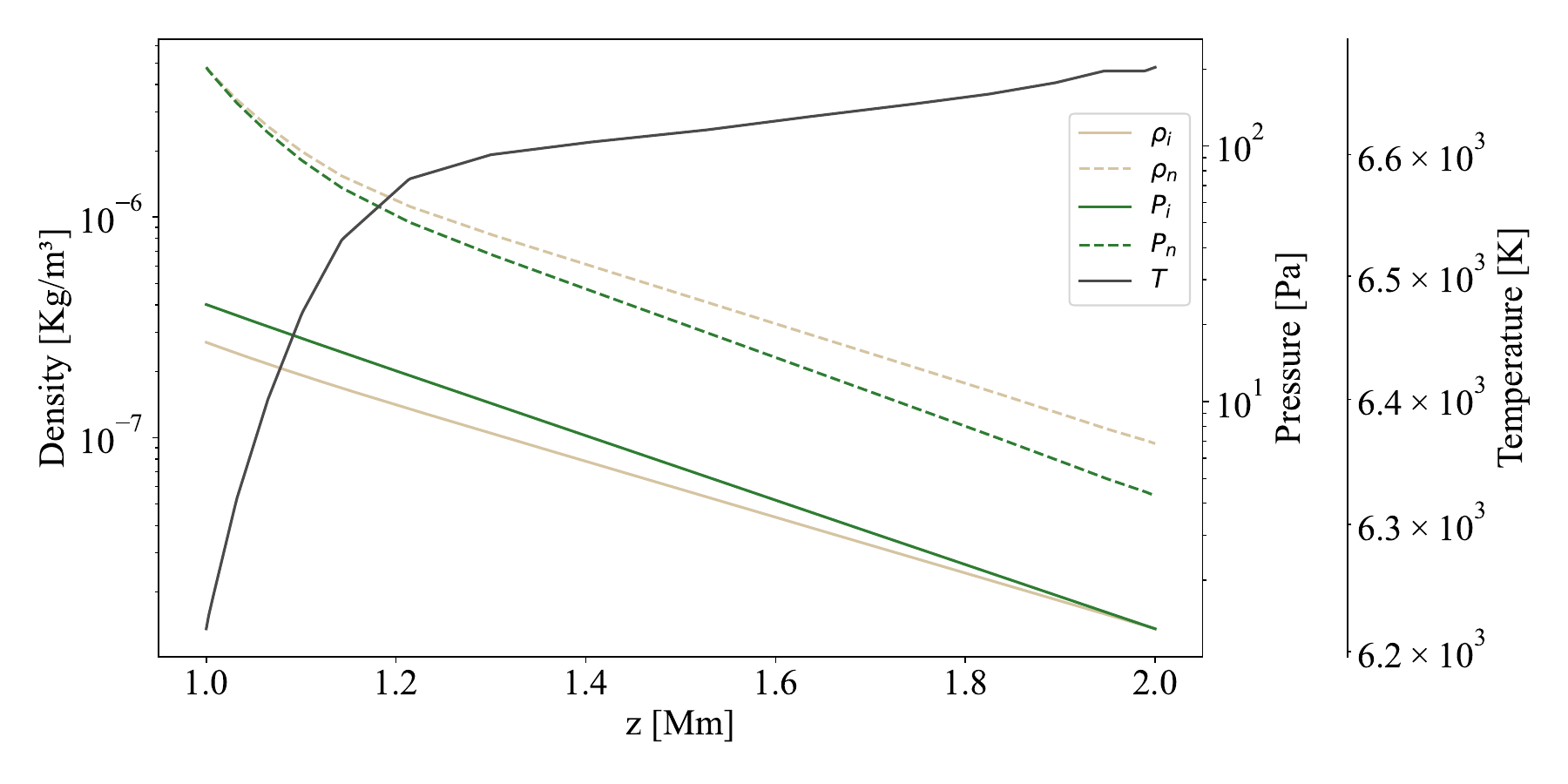}
    \caption{Close-up view of the atmospheric profiles within the computational domain, extending from 1 to 2~Mm in height. The interpolated temperature profile is shown by the black curve, with the corresponding values indicated by the color bar on the right. Additionally, we present the evolution of mass density (beige) and thermal pressure (green) for both the neutral (dashed lines) and charged (solid lines) fluid components.}
    \label{fig:scp}
\end{figure}

We investigate magnetic reconnection in the solar chromosphere for initial magnetic field strengths of $B_0$ = 100, 110, and 120~G. The computational domain extends from 1 to 2~Mm above the base of the photosphere and is centered at the origin of the $x$–$y$ plane, with dimensions $1\times 0.01 \times 1 ~\mathrm{Mm}^3$, representative of chromospheric conditions. For this purpose, we will employ a $2.5$ D system, in which the numerical grid will be $400 \times4 \times 400$ along the (x,y,z) axes. In this way, we obtain a spatial resolution of  $\Delta x = \Delta y = \Delta z = 2.5 \cdot 10^{-3}$ Mm in our mesh. Likewise, a value of $w_\eta = 0.066$~Mm is assumed for the width of the Ohmic resistivity. Charged species are assumed to be in hydrostatic equilibrium, while the neutral component satisfies ionization--recombination equilibrium. Gravitational effects are included self-consistently. Figure~\ref{fig:scp} presents a close-up view of the thermodynamic profiles corresponding to the chromospheric volume adopted in our simulations. This detailed perspective highlights that the background atmosphere is stratified. By employing the Avrett model to prescribe the temperature, pressure, and density profiles, the model self-consistently incorporates the effects of solar gravity. As a result, the simulated domain captures the sharp gradients and non-uniform conditions characteristic of the chromosphere, which are essential for accurately modeling collisional processes and magnetic reconnection in a partially ionized plasma.

Finally, it is necessary to specify the boundary conditions in accordance with the expected behavior of the solar atmosphere. For this reason, we employ a Dirichlet-type boundary condition. This condition implies that the function describing the boundary remains constant and equal to its initial value. This boundary condition is applied at the lateral boundaries of the simulation, since our computational domain is sufficiently large to encompass the principal dynamics of the system, which barely reach the edges of the numerical box. Similarly, this condition is imposed at the bottom boundary of the domain because the stratification of the solar chromosphere leads to an increase in density and pressure with depth, making it difficult for the magnetic reconnection dynamics to propagate downward and perturb the lower region. On the other hand, for the upper boundary of the domain, we implement a homogeneous Neumann-type boundary condition. This condition enables us to describe both the transmission of the reconnection dynamics into the less stratified upper layers and the ejection of chromospheric material into the corona.

\subsection{Morphology}

\begin{figure}
    \centering
    \hspace*{-0.2 cm}
    \includegraphics[width=1.0\linewidth]{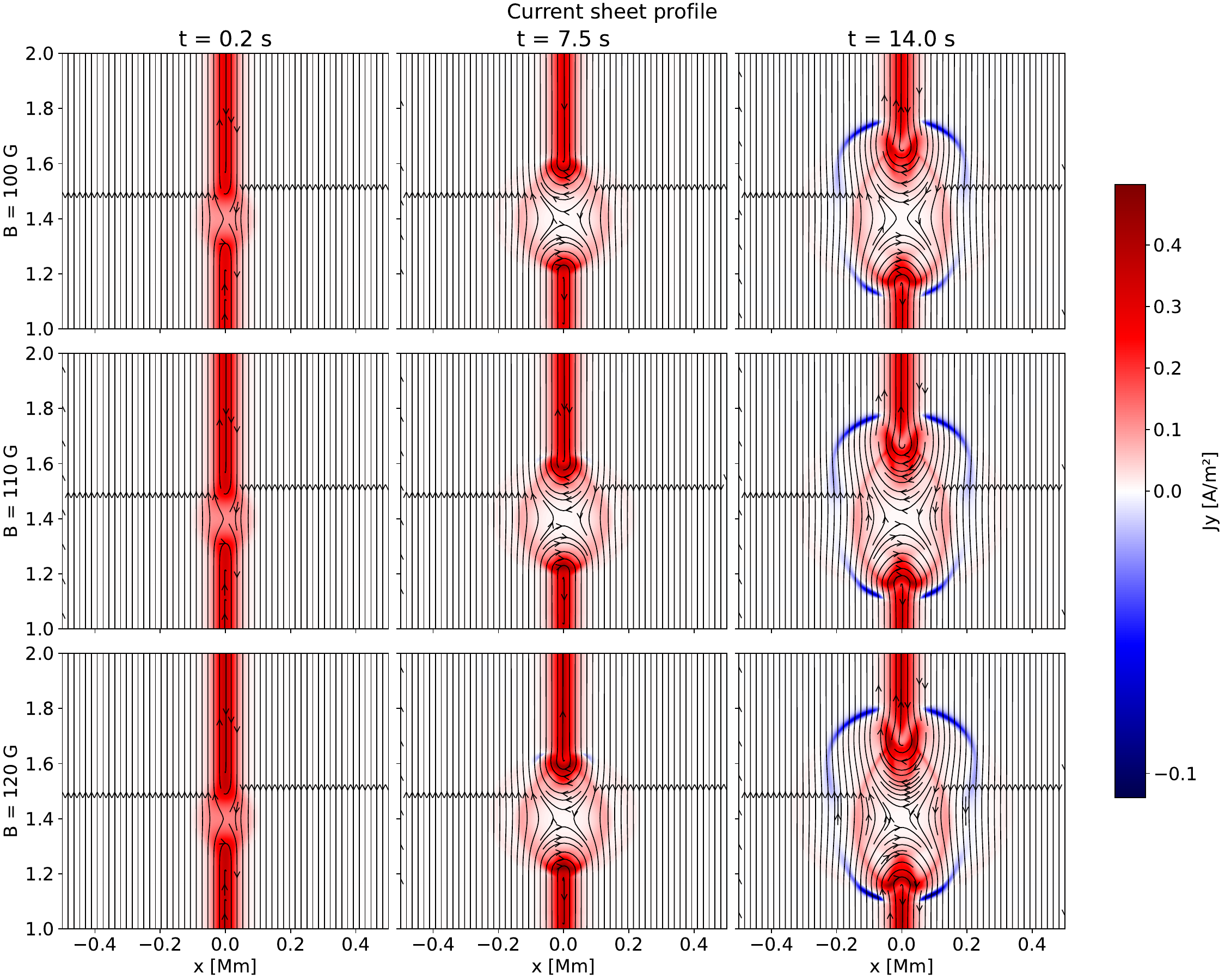}
    \caption{Evolution of the current density $J_y$ across different magnetic field strengths. Rows represent field strengths of 100, 110, and 120~G (top to bottom), illustrating the morphological response to increasing field intensity. Columns from left to right correspond to snapshots at $t = 0.2$, $7.5$, and $14.0$~s. On each current density panel, line contours indicate the flux  of the magnetic field.}
    \label{fig:JyME}
\end{figure}

First, we analyze the morphological evolution of the thermal, kinetic, and magnetic fields alongside the current density distribution. This analysis facilitates the identification of solar magnetic structures produced by magnetic reconnection within the chromosphere and allows us to quantify the impact of magnetic field strength on plasma heating and morphological development. Initially, we focus on the $J_y$ component of the current density, as it tracks the current flow across the polarity inversion line. The emergence of a current sheet, a primary diagnostic of ongoing magnetic reconnection, is monitored through this component \cite{li2018spectroscopic,reva2022observations}. To illustrate this, Figure \ref{fig:JyME} displays the $J_y$ profiles in the $xz$-plane at the mid-plane of the computational domain ($y=0$). The rows correspond to magnetic field strengths of 100, 110, and 120~G, respectively. The columns, from left to right, illustrate the temporal evolution at $t = 0.2$, $7.5$, and $14.0$~s. Finally, on each current density representation, line contours are shown, which indicate the flux and direction of the magnetic field throughout the evolution.

\begin{figure}
    \centering
    \hspace*{-0.2 cm}
    \includegraphics[width=1.0\linewidth]{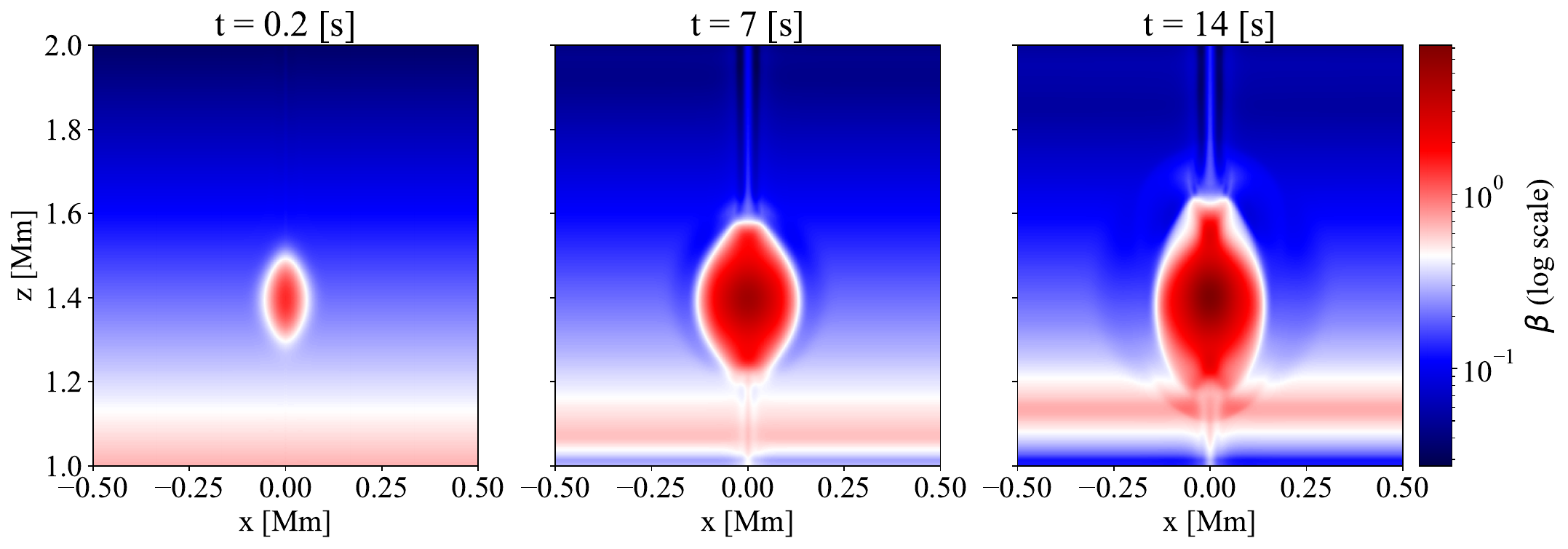}
    \caption{Morphological evolution of the plasma $\beta$ parameter at times $t=0.2$, 7, and 14 seconds. Although the initial magnetic field strength varies across the three cases, we display only the case $B0=100$~G for clarity. The evolution of $\beta$ is qualitatively similar for all three cases, with the largest increase occurring in the reconnection region.}
    \label{fig:Mbp}
\end{figure}

Initially, in Figure \ref{fig:JyME}, we observe that the thin current sheet structure typically expected in reconnection events does not manifest. Instead, following the onset of reconnection, the initial current density bifurcates into two distinct regions and undergoes a localized broadening around the diffusion region. This can be seen by observing the regions above and below the reconnection zone, where a significant and localized increase in current density can be observed, as well as the manifestation of closed loops in the magnetic field lines around these regions. This morphological divergence is attributed to the elevated role of thermal gas pressure in the chromospheric plasma; unlike the magnetically dominated (low-$\beta$) environment of the corona, the chromosphere exhibits higher plasma-$\beta$ conditions where fluid pressure significantly influences the reconnection dynamics. In fact, to observe this in more detail, Figure \ref{fig:Mbp} shows the morphological evolution of the plasma beta parameter, where its magnitude can be observed at times $t$ = 0.2, 7, and 14 seconds. In this case, although for each case we have a change in the initial magnetic field and, consequently, in the initial value of the plasma beta parameter, the morphological evolution of the plasma beta is very similar for all three cases, as is the case for the current density morphology; therefore, this figure is representative of all three cases. As can be seen, during the evolution of the reconnection, the magnitude of the plasma beta parameter increases, with its largest growth occurring in the region where reconnection takes place. Thus, in this region, the kinetic pressure exceeds the magnetic pressure, which is consistent with the fact that kinetic pressure becomes significant for the morphological evolution and the formation of structures. Despite this broadening, in Figure \ref{fig:JyME} the lower region displays an arcade-like structure through which the plasma circulates, while the upper region exhibits a characteristic thinning of the current flow, both of which are hallmarks of flare-associated reconnection \cite{reeves2022window,fletcher2024solar}. Furthermore, variations in the initial magnetic field strength do not qualitatively alter the global morphology; the primary effect of an increased field is a marginal enhancement of the peak current density.

\begin{figure}
    \centering
    \hspace*{-0.2 cm}
    \includegraphics[width=1.0\linewidth]{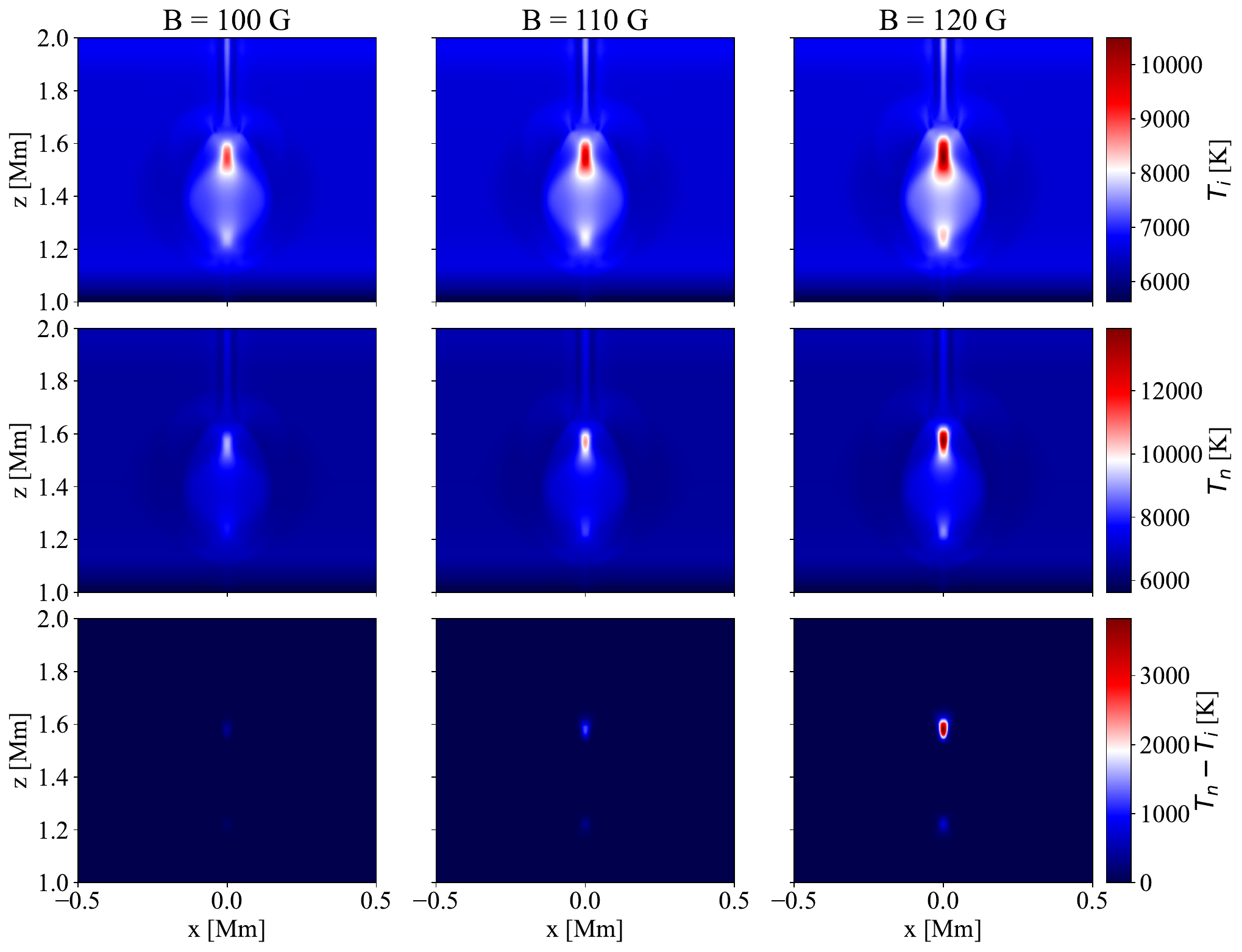}
    \caption{Temperature distributions at $14.0$~s for the charged (top row) and neutral (middle row) species, and the difference between them (bottom row), for initial magnetic field strengths $B_0 = 100$, $110$, and $120$~G (from left to right). The figure illustrates the morphological differences in chromospheric heating associated with magnetic reconnection and how variations in the initial field strength affect the thermal signatures of both species.}
    \label{fig:TME}
\end{figure}

Next, we examine the morphology of the temperature profiles of the charged and neutral species to characterize chromospheric heating from reconnection. This analysis quantifies the efficiency of magnetic‑to‑thermal energy conversion and assesses how variations in the initial magnetic field strength affect the thermal signatures. Figure~\ref{fig:TME} shows the temperature distributions for the charged (top row) and neutral (middle row) species at $14.0$~s, along with their difference (bottom row), for $B_0 = 100$, $110$, and $120$~G (left to right).

In the Figure \ref{fig:TME}, the first thing that can be observed is that, for both the charged and neutral species, the thermal evolution behaves similarly to the current density distribution; After reconnection occurs, the temperature bifurcates into two regions of maximum heating, one upper and one lower. These regions of localized heating coincide with the sites of maximum energy dissipation during the reconnection event. Physically, these features are driven by the bulk acceleration of plasma; the lower region manifests as a heating signature within the magnetic arcade, while the upper region corresponds to the upward ejection of a chromospheric lobe. On the other hand, it can be seen that, although the temperature morphology is similar for both fluids, the heating process is more efficient in the neutral species, with a maximum temperature gradient between the two fluids occurring in the regions located above and below the reconnection zone mentioned earlier.

\begin{figure}
    \centering
    \hspace*{-0.2 cm}
    \includegraphics[width=1.0\linewidth]{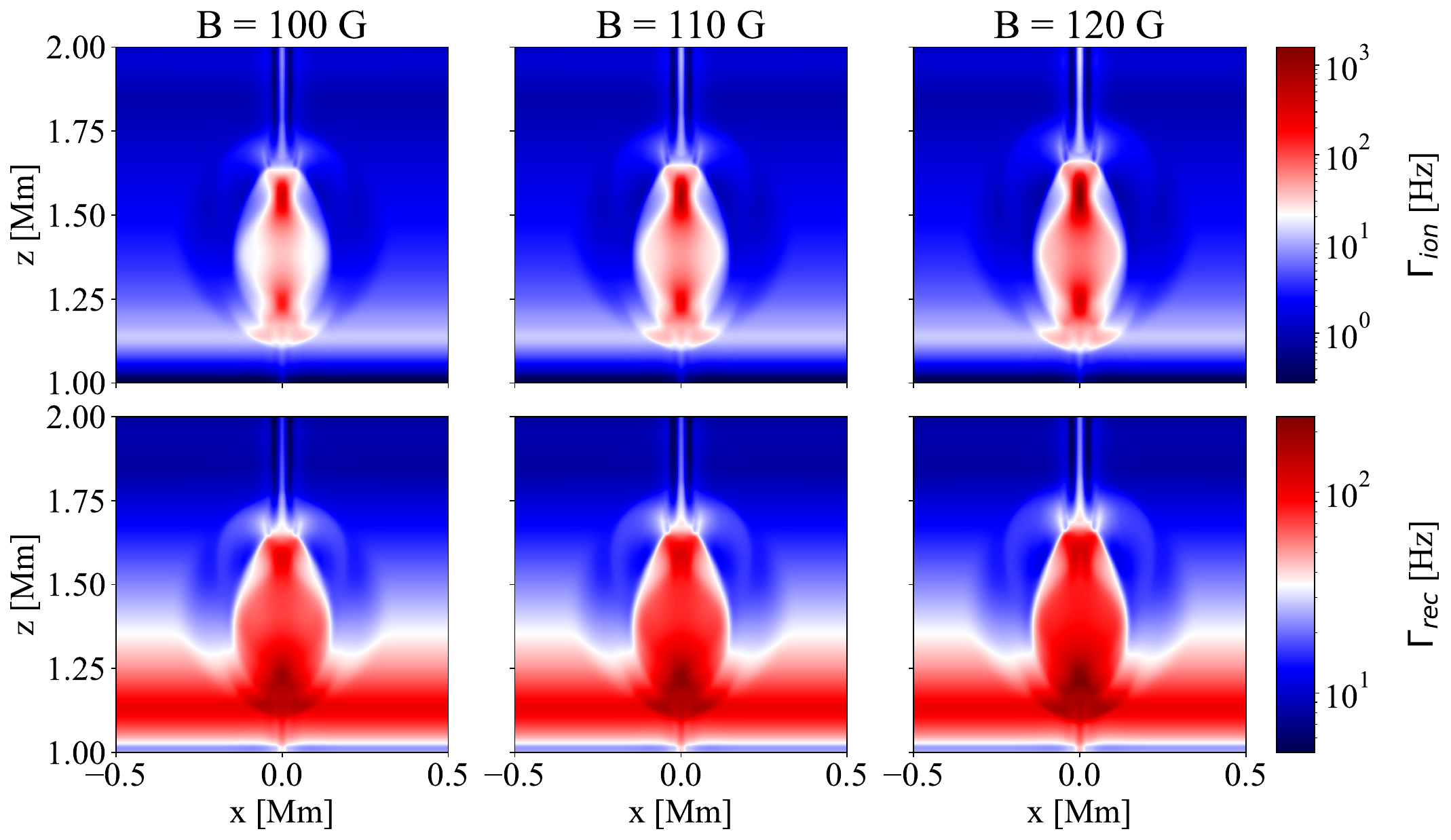}
    \caption{Ionization (top row) and recombination (bottom row) frequencies at time $t = 14$ seconds for initial magnetic field strengths $B_0$ = 100, 110, and 120 G, ordered from left to right. The figure illustrates the morphology of both frequencies following the reconnection event.}
    \label{fig:MIR}
\end{figure}

Lastly, we analyze the morphology of the ionization and recombination frequencies. To this end, Figure \ref{fig:MIR} shows the ionization (top row) and recombination (bottom row) frequencies at time $t = 14$ seconds. The frequencies are displayed for the cases $B_0$ = 100, 110, and 120 G, arranged from left to right, respectively.

In Figure \ref{fig:MIR} we can see that, as a result of reconnection, the ionization and recombination frequencies increased considerably due to the increase in species temperature. Additionally, although initially in our system the recombination frequencies in the chromosphere were higher than the ionization frequencies, after reconnection the ionization frequencies become dominant, exceeding the recombination frequencies, mainly near the plasma regions where the most significant temperature increases were recorded which are associated with the arcade and the ejected lobe of material.

\begin{figure}
    \centering
    \includegraphics[width=0.8\linewidth]{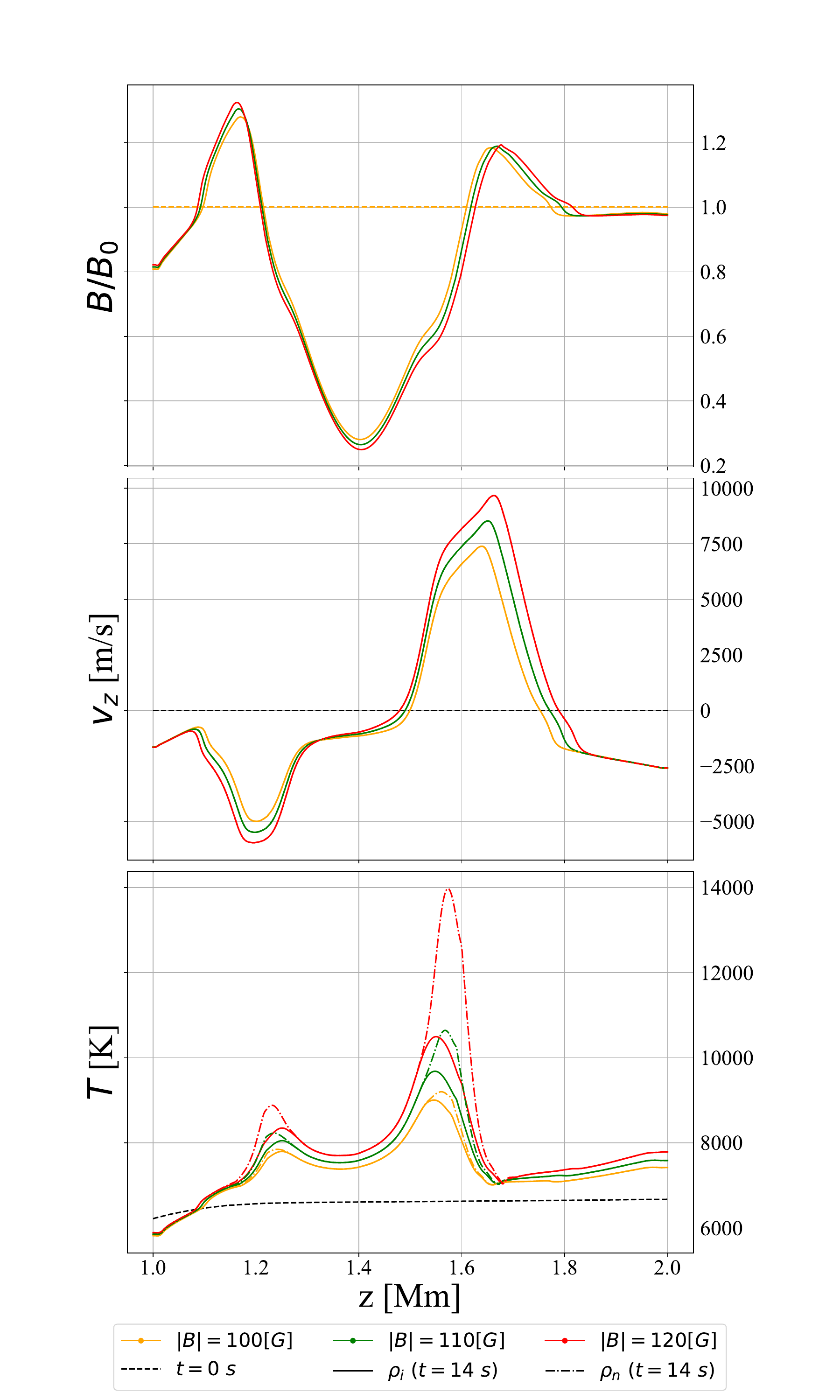}
    \caption{Normalized magnetic field magnitude (top), vertical profiles of the $z$-velocity component (center), and plasma temperature (bottom) as a function of height. Measurements are recorded along the $z$-axis at the origin of the $xy$-plane ($x=0, y=0$) within the chromospheric domain. Dashed lines represent the initial state ($t = 0$~s), while solid and dash-dotted lines correspond to the evolved state at $t = 14$~s for the charged and neutral species, respectively. The configurations for $B = 100$~G, $110$~G, and $120$~G are indicated by yellow, green, and red profiles, respectively.}
    \label{fig:EoVMFaT}
\end{figure}

To facilitate a quantitative assessment of the underlying dynamics, Figure \ref{fig:EoVMFaT} presents the normalized vertical profile of the magnetic field, as well as the $z$-velocity component and the plasma temperature for charged (solid lines) and neutral species (dash-dotted lines), measured along the polarity inversion line. These profiles compare the initial state ($t = 0$~s, dashed lines) with the evolved state at $t = 14$~s for the $100$~G (yellow), $110$~G (green), and $120$~G (red) configurations.

In accordance with the magnetic evolution, shown in the top panel of the figure \ref{fig:EoVMFaT}, a rapid depletion of the magnetic field occurs within the diffusion region centered at the reconnection point. Conversely, two distinct zones of magnetic compression emerge above and below this site, where the field strength increases significantly relative to the initial state. While the initial field magnitudes were $100$, $110$, and $120$~G (representing $10\%$ and $20\%$ increments relative to the baseline case), the system evolves over $14$~s to reach peak intensities of $127.90$, $143.41$, and $158.89$~G in the lower region at a height of $1.16$~Mm. In the upper region, peak values of $118.37$, $130.75$, and $143.08$~G are recorded at $1.67$~Mm. These altitudes coincide with the structural maxima observed in the $J_y$ morphological distribution. Quantitatively, the lower region exhibits field enhancements of $27.9\%$, $30.37\%$, and $32.41\%$ for the $100$, $110$, and $120$~G configurations, respectively, while the upper region shows increases of $18.37\%$, $18.86\%$, and $19.23\%$. These results indicate that a higher initial magnetic flux density leads to a greater percentage increase in post-reconnection field maxima, suggesting a more efficient magnetic flux processing, particularly in the lower region. This behavior is linked to the localized decrease in magnetic pressure relative to the plasma kinetic pressure. Furthermore, the lower percentage increase in the upper region suggests that energy release and particle acceleration are more dominant there, resulting in a higher conversion rate of stored magnetic energy into kinetic and thermal forms.

Regarding the vertical velocity ($v_z$), the middle panel illustrates the development of bidirectional plasma outflows emanating from the reconnection site for both the charged and neutral species. It is important to note that the maximum velocity differential between the neutral and charged species is negligible, residing on the order of $10^{-1}$~m/s. For this reason, although the velocity profiles for both the neutral and charged species are shown, they overlap, and only the profile of the charged species is displayed. Nevertheless, given that the velocities are nearly identical for the phenomenon, the dynamical evolution of the velocity field and the associated growth percentages are effectively identical for both fluid populations. Given this strong coupling, a unified description is applied to both species throughout this analysis. At $t = 14$~s, the downward-propagating pulses reach maximum velocities of $4988.44$, $5479.33$, and $5948.27$~m/s for the $100$, $110$, and $120$~G cases, respectively, at an altitude of $z = 1.2$~Mm. Conversely, the upward-directed pulses attain significantly higher velocities of $7387.00$, $8527.01$, and $9670.26$~m/s, peaking at $z = 1.65$~Mm. 

These results demonstrate that plasma acceleration scales positively with the initial magnetic field intensity; specifically, the $110$ and $120$~G configurations exhibit velocity increments of $9.84\%$ and $19.24\%$ in the lower pulse, and $15.43\%$ and $30.90\%$ in the upper pulse, relative to the $100$~G baseline. This asymmetry suggests that acceleration is markedly more efficient in the upper region. This disparity is attributed to atmospheric stratification: the downward pulse is decelerated by the increasing density and pressure gradients in the lower chromosphere, whereas the upward pulse propagates into a lower-density environment, facilitating rapid acceleration. Ultimately, these kinematic signatures corroborate our morphological findings, identifying the lower pulse as plasma driven into the magnetic arcade and the upper pulse as a chromospheric jet ejected toward the solar corona. 

Finally, as it is shown in the bottom panel of the figure \ref{fig:EoVMFaT}, an analysis of the thermal profiles along the polarity inversion line reveals that the magnitude of chromospheric heating scales positively with the initial magnetic field strength. For the charged species, we identify two distinct thermal maxima corresponding to the previously discussed regions. In the lower region ($z = 1.25$~Mm), peak temperatures reach $7798.14$, $8050.89$, and $8349.29$~K for the $100$, $110$, and $120$~G cases, respectively. In the upper region ($z = 1.55$~Mm), the temperature maxima are significantly higher, reaching $9006$, $9682$, and $10495$~K across the same magnetic configurations. Given that the initial temperatures at $z = 1.25$~Mm and $z = 1.55$~Mm were $6581.41$~K and $6623.09$~K, respectively, these results represent thermal enhancements ranging from $18.49\%$ to $58.46\%$. This significant temperature rise is indicative of the enhanced energy-dissipation efficiency associated with higher magnetic flux densities, particularly in the upward-ejected chromospheric material. On the other hand, the behavior of the temperature is quite similar to that of the charged species, with an upper and a lower thermal maximum, whose locations are the same as those of the charged species but slightly displaced. However, for the lower region, the peak temperatures reach values of $7848.22$~K, $8230.97$~K, and $8882.82$~K, while in the upper region the temperature reaches values of $9201.31$~K, $10644.54$~K, and $13975.88$~K. Given that the neutral species share the same initial temperature as the charged species, these results represent thermal enhancements ranging from $19.25\%$ to $111.02\%$. This shows that the thermal energy released in the neutral species during magnetic reconnection is particularly significant, possibly because the neutral species is dominated by kinetic and thermal processes, allowing the system to reach higher temperatures.

\subsection{Collision Terms Impact}

Continuing the investigation of the reconnection dynamics, and given that a primary objective of this study is to characterize reconnection within a partially ionized environment, we examine the multi-species coupling between the charged and neutral components. Specifically, we evaluate the impact of elastic and inelastic collisional processes, alongside the influence of the magnetic field on the spatial and temporal evolution of ionization and recombination rates throughout the reconnection event.

\begin{figure}
    \centering
    \hspace*{-0.2 cm}
    \includegraphics[width=1.0\linewidth]{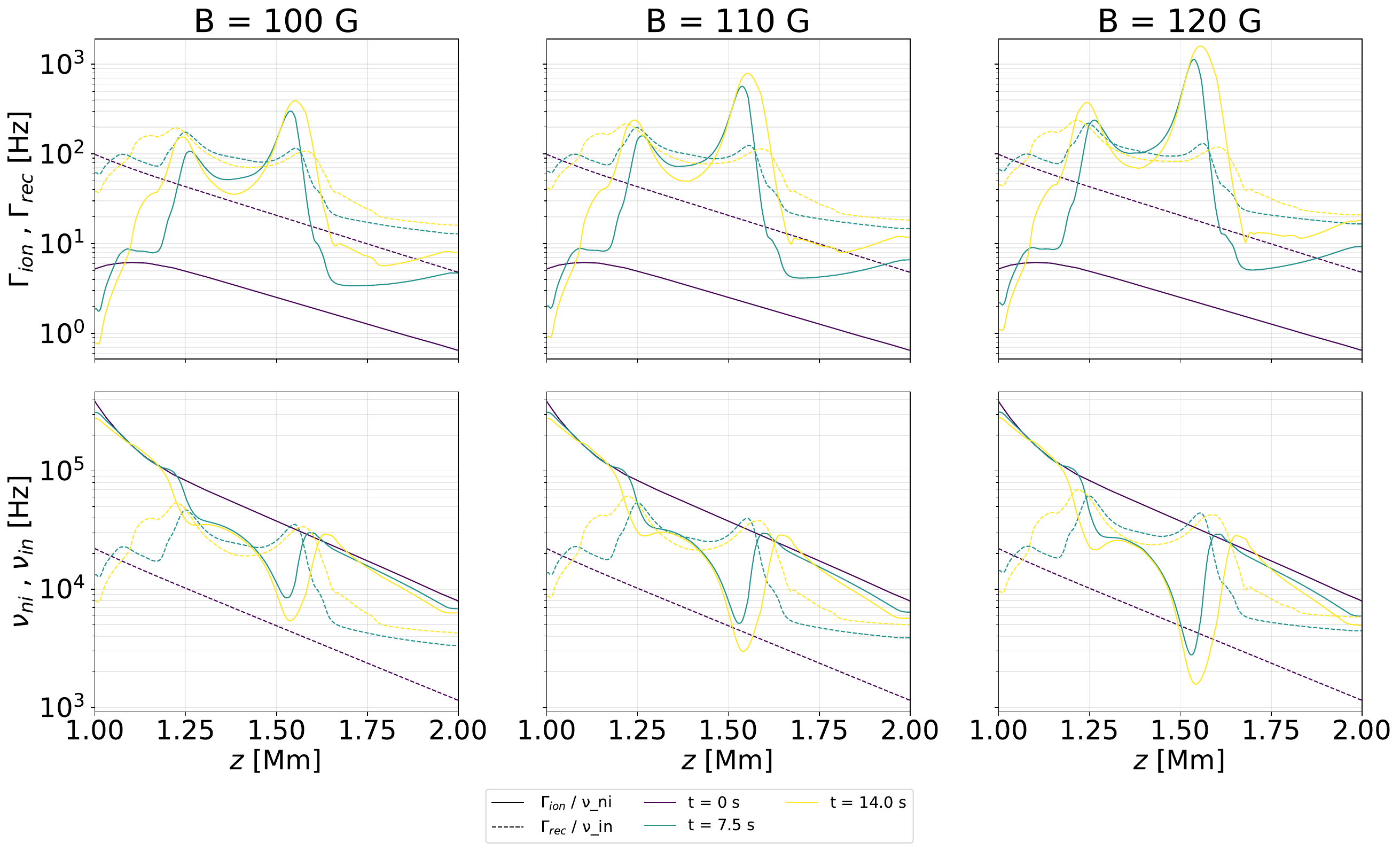}
    \caption{Evolution of collisionality during the reconnection process. The rows display the inelastic frequencies of ionization (solid) and recombination (dashed), and elastic frequencies $\nu_{ni}$ (solid) and $\nu_{in}$ (dashed), while columns correspond to the initial magnetic field configurations of $100$, $110$, and $120$~G. Frequencies are plotted as a function of height $z$, extracted along the vertical axis at the center of the current sheet. The temporal evolution is represented at $t = 0$~s (purple), $t = 7.5$~s (blue), and $t = 14$~s (yellow).}
    \label{fig:IRF}
\end{figure}

To this end, Figure \ref{fig:IRF} illustrates the evolution of the inelastic collisionality, specifically the ionization (solid) and recombination (dashed) rates, alongside the elastic collision frequencies, $\nu_{ni}$ (solid) and $\nu_{in}$ (dashed). The rows categorize the specific frequency under analysis, while the columns correspond to the varying initial magnetic field intensities. These frequencies are calculated as a function of altitude $z$ and are sampled along the vertical axis at the center of the current sheet. The temporal evolution is captured at three discrete intervals: the initial state ($t = 0$~s, purple), $t = 7.5$~s (blue), and $t = 14$~s (yellow).

First, we examine the inelastic collisional processes. As the system evolves, the ionization and recombination rates increase significantly, particularly in regions experiencing peak thermal enhancement. At $t = 14$~s, we analyze the maximum inelastic frequencies and their corresponding altitudes. In the lower region, ionization frequencies reach $154.92$, $239.34$, and $374.68$~Hz, while recombination frequencies are $196.75$, $218.10$, and $238.11$~Hz for the $100$, $110$, and $120$~G configurations, respectively. These ionization maxima are localized at $z = 1.24$~Mm, with recombination peaking at $z = 1.22$~Mm. For comparison, the initial ionization and recombination frequencies in this region were $5.06$~Hz and $47.21$~Hz, respectively. 

In the upper region ionization frequencies attain $391.02$, $790.39$, and $1598.78$ Hz, while recombination rates are $107.48$, $114.38$, and $119.12$~Hz. The ionization maxima occur at $z = 1.55$~Mm, with recombination at $z = 1.58$~Mm. Initially, these rates were $2.19$~Hz (ionization) and $16.37$~Hz (recombination). While both rates scale with time and magnetic field intensity, the ionization frequency exhibits a vastly superior percentage increase, eventually reaching or surpassing recombination scales. This trend highlights enhanced ionization activity and a substantial increase in charged-neutral coupling. These findings align with observational signatures of ionization traces and enhanced X-ray emissions recorded at the base and cusp of flare arcades, as well as within ejected lobes \cite{fletcher2024solar,vievering2021foxsi}.

Next, we analyze the extrema of the elastic collision frequencies in the same instant $t=14$~s, specifically focusing on the $\nu_{in}$ maxima. In the lower region, peak $\nu_{in}$ values of $53801.96$, $61302.52$, and $68955.15$~Hz are recorded for the $100$, $110$, and $120$~G cases, respectively, at an altitude of $z = 1.27$~Mm. Conversely, the upper region exhibits lower maxima of $33621.86$, $37751.42$, and $42461.17$~Hz at $z = 1.58$~Mm. These results demonstrate that both the absolute magnitude and the relative increase of $\nu_{in}$ are significantly more pronounced in the lower chromospheric layers. This trend is physically consistent with the plasma environment: although the upper region experiences higher thermal enhancements, the ion-neutral collision frequency $\nu_{in}$ is primarily governed by the charged-species density, which remains dominant in the more stratified lower region.

In contrast, at $t=14$~s the neutral-ion collision frequency, $\nu_{ni}$, exhibits a significant decrease during the reconnection event. In the lower region ($z = 1.27$~Mm), we record minima of $34791.47$, $28168.86$, and $21482.51$~Hz for the $100$, $110$, and $120$~G cases, respectively. This depletion is even more pronounced in the upper region ($z = 1.54$~Mm), where frequencies drop to $5405.97$, $2976.06$, and $1566.56$~Hz. Although the temperature rise might suggest an increase in collisionality, $\nu_{ni}$ is fundamentally constrained by the local neutral density as it is shown in equations \ref{eq:Alpha} and \ref{eq:NiNn}. As demonstrated by the elevated ionization rates discussed previously, the neutral population is rapidly depleted as neutrals are converted into charged species. This localized reduction in neutral density outweighs the thermal effects, leading to the observed decline in $\nu_{ni}$. Consequently, the most substantial decrease occurs in the upper region, where the ionization efficiency and the resulting neutral species depletion are most significant. 

\begin{figure}
    \centering
    \includegraphics[width=0.9\linewidth]{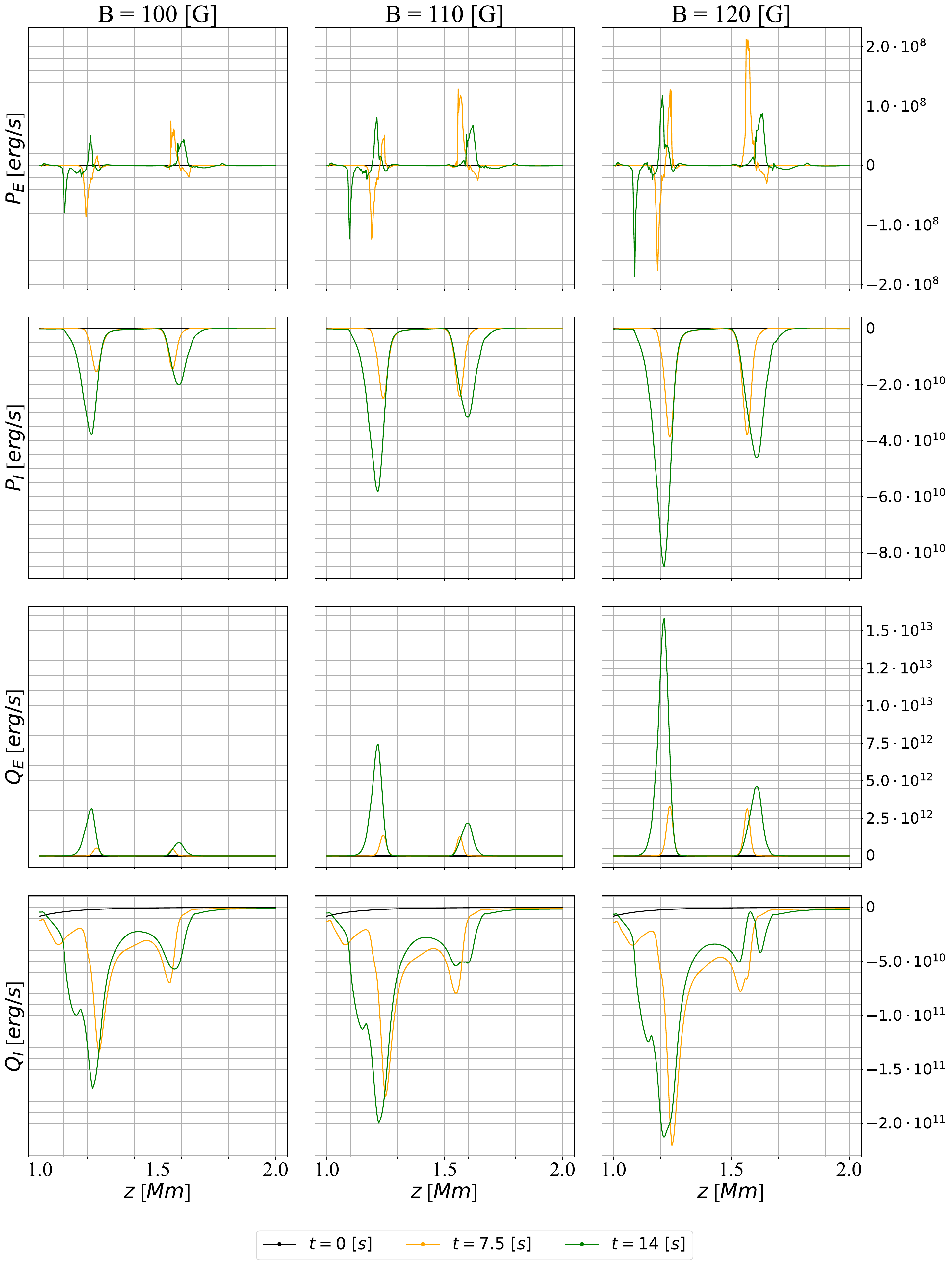}
    \caption{Temporal evolution of the collisional energy transfer terms during the reconnection process. The rows categorize the components of the interaction term $H_i$, representing the kinetic power transfer via elastic ($P_E$) and inelastic ($P_I$) collisions, alongside the inter-species heat exchange for elastic ($Q_E$) and inelastic ($Q_I$) processes. Columns denote the initial magnetic field configurations of $100$, $110$, and $120$~G. Each term is plotted as a function of altitude $z$, extracted along the vertical axis at the center of the current sheet. The profiles compare the initial state ($t = 0$~s, black) with the evolved states at $t = 7.5$~s (yellow) and $t = 14$~s (green).}
    \label{fig:EoP}
\end{figure}

Finally, we characterize the impact of collisional processes on the energetic evolution of the system. To this end, we analyze the individual components of the collisional interaction term $H_i$, categorized as follows:
\begin{eqnarray}
    P_E &=& \frac{1}{2} \alpha \left(u_n^2 - u_i^2\right), \label{eq:PE} \\
    P_I &=& \frac{1}{2} \left(\rho_{n} u_n^2 \Gamma^{\mathrm{ion}} - \rho_{i} u_i^2 \Gamma^{\mathrm{rec}}\right), \label{eq:PI} \\
    Q_E &=& \frac{\alpha}{\gamma - 1} \frac{k_B}{\bar{m}} \left(T_n - T_i \right), \label{eq:QE} \\
    Q_I &=& \frac{1}{\gamma - 1} \frac{k_B}{\bar{m}} \left( \rho_{n} T_n \Gamma^{\mathrm{ion}} - \rho_{i} T_i \Gamma^{\mathrm{rec}} \right). \label{eq:QI}
\end{eqnarray}
In this framework, $P_E$ and $P_I$ represent the kinetic power transfer driven by elastic and inelastic collisions, respectively, while $Q_E$ and $Q_I$ denote the inter-species heat exchange. Figure \ref{fig:EoP} illustrates the temporal evolution of these energy transfer rates. The rows correspond to the individual components of $H_i$, and the columns categorize the results by initial magnetic field strength. Consistent with the previous analysis of collision frequencies, these energetic terms are plotted as a function of altitude $z$ along the mid-plane of the domain, comparing the initial state ($t = 0$~s, black) with the evolved states at $t = 7.5$~s (yellow) and $t = 14$~s (green).

Regarding the kinetic power transfer terms, $P_E$ and $P_I$, their impact on the plasma dynamics is negligible compared to both the total energy release and the thermal exchange rates. This result suggests a regime of strong species coupling; as the ion and neutral fluids move nearly in unison, the relative drift velocity remains minimal, effectively nullifying the collisional power terms. In contrast, the elastic heat transfer term, $Q_E$, represents the dominant contribution to the energetic evolution. This indicates that elastic collisions, primarily through inter-species friction, drive a substantial transfer of thermal energy, contributing significantly to the heating of the charged species. Furthermore, the magnitude of $Q_E$ scales positively with the initial magnetic field intensity, corroborating our previous observations of enhanced chromospheric heating at higher flux densities. 

Finally, the inelastic heat transfer term, $Q_I$, provides the second-largest contribution to the energy budget. Unlike $Q_E$, the $Q_I$ term represents a thermal energy flux from the charged species to the neutral background. This is physically consistent with the reconnection dynamics: the charged species is preferentially heated by magnetic dissipation and elastic interactions, and subsequently transfers a portion of this thermal energy to the neutral population during recombination events.

\subsection{Reconnection rates}

A critical parameter in this study is the reconnection rate, which dictates the timescale of the magnetic flux processing and determines the overall efficiency of the reconnection event \cite{birn2007reconnection}. The dimensionless reconnection rate, often expressed as the Alfv\'enic Mach number $M_A$, is defined as:
\begin{equation}
    M_A = \frac{u_{in}}{u_A}, \quad  \quad u_A = \frac{|\mathbf{B}|}{\sqrt{\mu_0 \rho}}.
\end{equation}
In this framework, $u_{in}$ denotes the plasma inflow velocity into the diffusion region (corresponding here to the $u_x$ component), while $u_A$ represents the local Alfv\'en speed. Typically, fast reconnection rates in solar environments are found within the range of $0.01 \lesssim M_A \lesssim 0.1$ \cite{shibata2011solar}. However, these values are not strictly bounded; regimes of "fast" or "explosive" reconnection can surpass these scales depending on the local plasma-$\beta$ and the specific resistivity model employed \cite{landinez2024systematic}.

\begin{figure}
    \centering
    \hspace*{-0.2 cm}
    \includegraphics[width=0.8\linewidth]{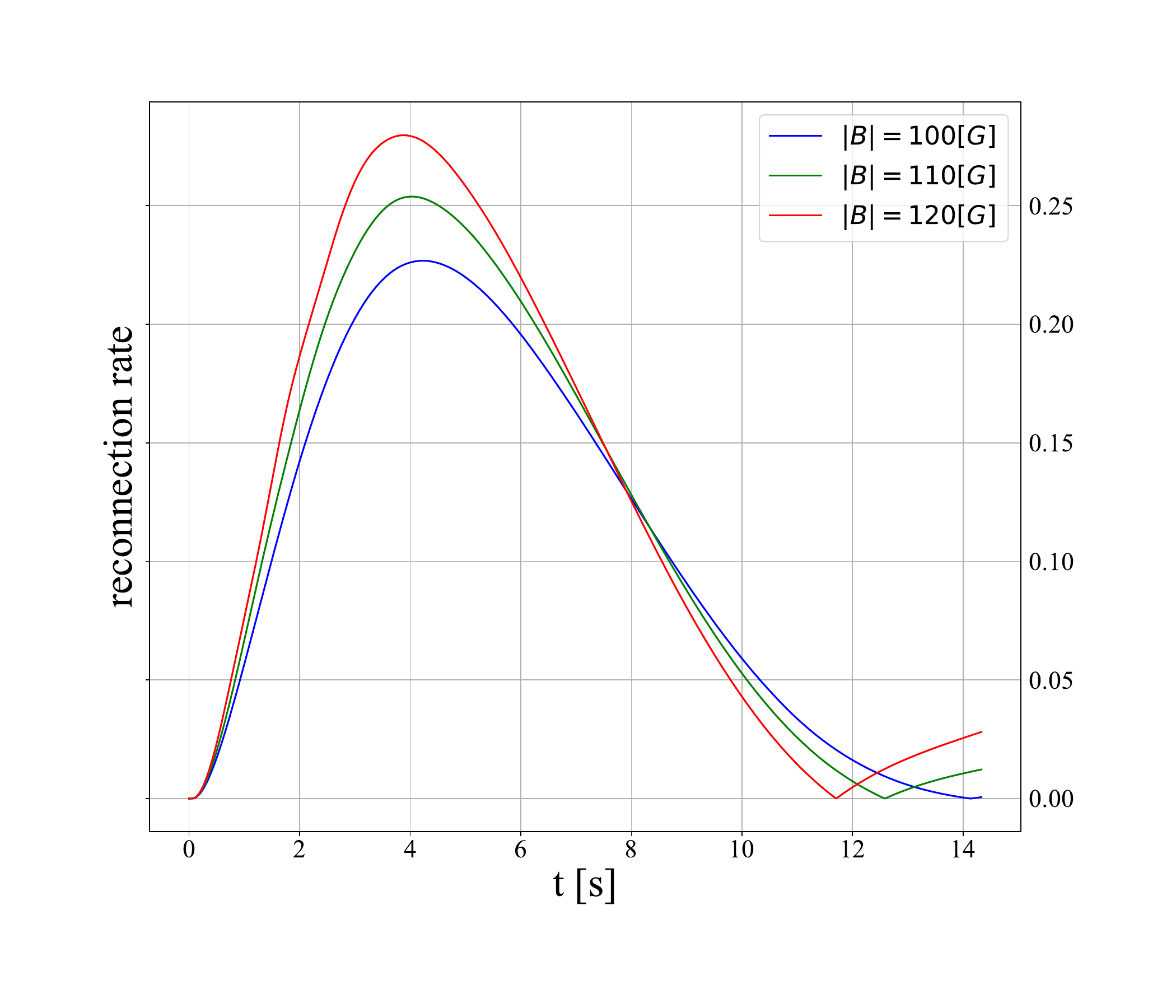}
    \caption{Temporal evolution of the magnetic reconnection rate $M_A$. Profiles correspond to magnetic field strengths of 100~G (blue), 110~G (green), and 120~G (red). Measurements were recorded at the edge of the current sheet at the reconnection site, $(x,y,z) = (0.2, 0, 1.4)$~[Mm], over a 14-second interval following onset.}
    \label{fig:RR}
\end{figure}

To this end, Figure \ref{fig:RR} illustrates the temporal evolution of the reconnection rates for the $100$~G (blue), $110$~G (green), and $120$~G (red) configurations. The dynamics are captured over a $14$-second interval, during which the reconnection rate accelerates to a peak value, denoting the phase of maximum flux processing, before subsequently decaying as the available magnetic energy is depleted.

At first glance, the temporal interval required to reach the onset of peak reconnection exhibits minimal observational variance. However, a precise recording of the peak timescales reveals values of $4.2285$, $4.0265$, and $3.878$~s for the $100$, $110$, and $120$~G configurations, respectively. These results demonstrate that the onset of the maximum reconnection rate scales inversely with the initial magnetic field strength. While the dimensionless rate $M_A$ typically decays following the primary maximum, the observed secondary enhancement could be attributed to the fact that the neutral species was not in thermal equilibrium. However, previous analyses have shown that this does not significantly perturb the overall dynamics of the problem. Consequently, the secondary enhancement of the reconnection process is likely a consequence of the intrinsic system dynamics, demonstrating that the system remains capable of undergoing further energy release.

In contrast, the peak reconnection rates are $0.226$, $0.253$, and $0.279$ for the $100$, $110$, and $120$~G configurations, respectively, demonstrating a positive correlation with the initial magnetic field intensity. While typical solar flare reconnection rates are reported in the range $0.01 \lesssim M_A \lesssim 0.1$, our results, though slightly exceeding this interval, remain within the expected order of magnitude. This discrepancy is primarily attributed to the simulation domain size, which is significantly smaller than the characteristic length scales of a typical solar flare. 

To contextualize these values, we consider the maximum reconnection rate predicted by the Petschek model, $M_{A,max} = \pi / (8 \ln S)$ \cite{birn2007reconnection}. For our parameters, this analytical model yields a rate of $0.074$. Given that the Lundquist number is defined as $S = \mu_0 L u_A / \eta$ \cite{pontin2022magnetic}, where $L$ is the characteristic length of the reconnection site, the reconnection rate effectively exhibits an inverse relationship with the system's spatial scale. Consequently, at realistic solar scales, the maximum reconnection rates would likely converge toward standard observational values. Finally, it should be noted that the Petschek model is derived from 2D steady-state equilibrium conditions. In 3D geometries, the precise characterization of reconnection rates remains an area of active research, with current models suggesting that maximum rates can exceed classical 2D predictions \cite{pontin2022magnetic}.

\subsection{Energy Release}

Next, we quantify the energy release and partitioning during the reconnection event. Typical solar flares dissipate energy in the range of $10^{28}$ to $10^{32}$~erg \cite{shibata2011solar}, distributed primarily among kinetic, enthalpy, and magnetic energy reservoirs. To evaluate these components, we define the following energy densities:
\begin{equation}
    k_{\alpha} = \frac{1}{2} \rho_{\alpha} u_{\alpha}^2, \quad h_{\alpha} = P_{\alpha} + \rho_{\alpha} e_{\alpha}, \quad e_M = \frac{B^2}{2 \mu_0},
\end{equation}
where $k_{\alpha}$ represents the kinetic energy density for the species $\alpha$ (charged $i$, neutral $n$), $h_{\alpha}$ is the corresponding enthalpy density (with internal energy $e_{\alpha}$), and $e_M$ denotes the magnetic energy density. The total energy content within a specified chromospheric volume $V$ is obtained through volume integration:
\begin{eqnarray}
    K_{\alpha} &=& \int_V k_{\alpha} \, dV, \quad H_{\alpha} = \int_V h_{\alpha} \, dV, \quad E_M = \int_V e_{M} \, dV. \label{eq:Energies}
\end{eqnarray}
The net energy release is subsequently determined by the temporal variation of these global quantities relative to the initial state of the system ($t = 0$~s). 

Furthermore, we account for the irreversible energy dissipation driven by magnetic resistivity. The volumetric power density associated with Joule heating is defined as:
\begin{equation}
    e_\eta = \eta J^2,
\end{equation}
where $\eta$ corresponds to the plasma resistivity and $J$ is the current density. Integrating this density over the volume yields the instantaneous dissipated power, $E_\eta = \int_V e_\eta \, dV$. Finally, the cumulative dissipated energy is calculated by integrating $E_\eta$ over the full $15$-second temporal evolution.

\begin{figure}
    \centering
    \hspace*{-0.2 cm}
    \includegraphics[width=\linewidth]{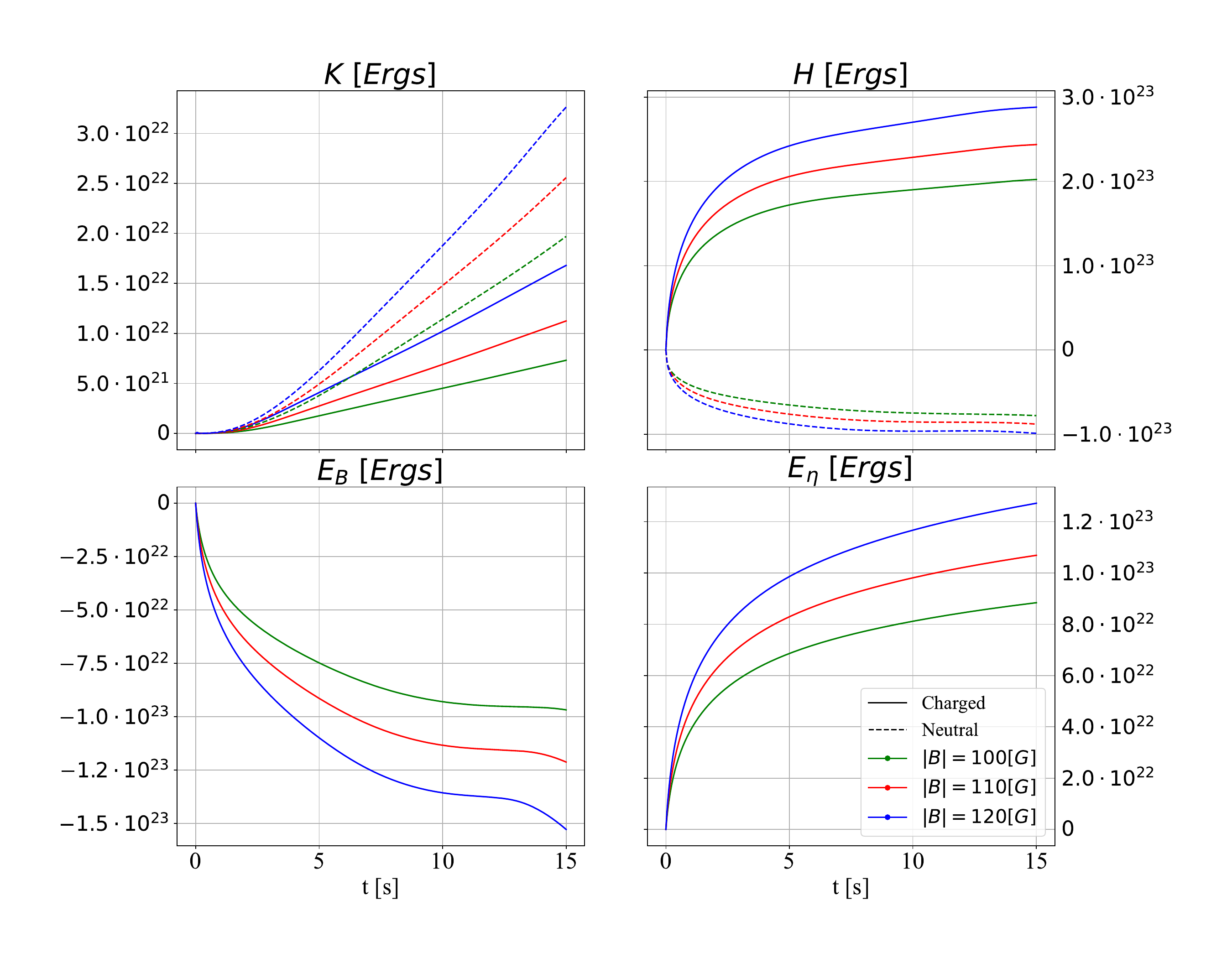}
    \caption{Temporal evolution of the energy components for the charged species (solid lines) and neutral species (dashed lines). The panels illustrate the variation in kinetic energy ($\Delta K$), enthalpy ($\Delta H$), magnetic energy ($\Delta E_M$), and cumulative Ohmic dissipation ($E_\eta$), calculated relative to the initial state. Energies are integrated over a chromospheric volume of $0.4 \times 0.01 \times 0.4$~Mm$^3$ centered in the $xy$-plane, spanning a $15$-second interval following the onset of reconnection. Results are categorized by initial magnetic field strengths: $100$~G (green), $110$~G (red), and $120$~G (blue).}
    \label{fig:LoE}
\end{figure}

In this study, the integration volume is defined by the dimensions $0.4 \times 0.01 \times 0.4$~Mm$^3$, spanning the altitude range $z \in [1.4, 1.8]$~Mm and centered at the origin of the $xy$-plane. To characterize the energetic evolution, Figure \ref{fig:LoE} illustrates the variation in kinetic energy ($\Delta K$) and enthalpy ($\Delta H$) for both charged (solid lines) and neutral species (dashed lines), alongside the fluctuations in magnetic energy ($\Delta E_M$) and cumulative Ohmic dissipation ($E_\eta$). These variations are calculated relative to the initial state over a $15$-second temporal evolution. 

As the system is initialized from a state of rest, the initial kinetic energies and Ohmic dissipation are identically zero. The baseline enthalpy values are $1.79 \times 10^{23}$~erg for the charged species and $6.55 \times 10^{23}$~erg for the neutral species across all configurations. Finally, the initial magnetic energy content is $6.48 \times 10^{23}$, $7.85 \times 10^{23}$, and $9.33 \times 10^{23}$~erg for the $100$, $110$, and $120$~G cases, respectively.

First, we analyze the variation in magnetic energy, which serves as the primary reservoir for energy extraction during the reconnection process. This variation is shown in the bottom-left panel in the figure \ref{fig:LoE}. At $t = 15$~s, the total released magnetic energy attains values of $9.67 \times 10^{22}$, $1.21 \times 10^{23}$, and $1.53 \times 10^{23}$~erg for the $100$, $110$, and $120$~G cases, respectively. While a linear scaling of the released energy percentage might be expected, given that only the initial field magnitude is varied, our results reveal a non-linear trend. Specifically, approximately $14.92\%$, $15.41\%$, and $16.40\%$ of the initial magnetic energy was liberated in each respective case. These findings indicate that the efficiency of magnetic energy conversion during reconnection scales positively with the initial magnetic field strength. 

Regarding the enthalpy evolution, in the upper-right panel of the figure \ref{fig:LoE}, we observe significant variations across both species at $t = 15$~s. For the charged species, enthalpy increases by $2.02 \times 10^{23}$, $2.43 \times 10^{23}$, and $2.88 \times 10^{23}$~erg, while the neutral species exhibits a corresponding decrease of $7.79 \times 10^{22}$, $2.55 \times 10^{22}$, and $3.26 \times 10^{22}$~erg for the $100$, $110$, and $120$~G cases, respectively. These gains in the charged-species enthalpy represent $112.84\%$, $135.63\%$, and $160.33\%$ of their baseline values. Conversely, the neutral population shows a relative decrease of $11.90\%$, $13.42\%$, and $15.11\%$. These results confirm that the magnitude of enthalpy flux scales directly with the initial magnetic field intensity. The enthalpy is a critical diagnostic in this framework, as it accounts for both internal energy changes and the work performed through expansion \cite{birn2009energy}. Our findings suggest that while magnetic reconnection drives both bulk acceleration and localized heating, the collisional coupling mediates a net thermal energy transfer from the neutral background to the charged fluid. This inter-species coupling effectively depletes the neutral enthalpy reservoir to drive the observed intensification of charged-species heating, highlighting the essential role of collisions in the energetic partitioning of the partially ionized plasma.

In the upper-left panel of the figure \ref{fig:LoE}, the kinetic energy values for the charged species at $t = 15$~s are $7.31 \times 10^{21}$, $1.12 \times 10^{22}$, and $1.68 \times 10^{22}$~erg, while the neutral species attain $1.96 \times 10^{22}$, $2.55 \times 10^{22}$, and $3.26 \times 10^{22}$~erg for the $100$, $110$, and $120$~G cases, respectively. Normalizing these values to the $100$~G baseline, the kinetic energy of the charged species increases by $53.21\%$ and $129.82\%$, whereas the neutral species shows gains of $29.85\%$ and $65.64\%$. Summing both components, the total kinetic energy growth relative to the $100$~G case is $36.34\%$ and $83.01\%$ for the $110$ and $120$~G configurations, respectively.

These results indicate that while the kinetic energy of the neutral population remains higher in absolute terms, owing to its significantly larger mass density, the charged species exhibits a much more rapid rate of acceleration. This discrepancy is physically rooted in the collisional coupling: as density increases, the neutral population acts as a momentum reservoir, transferring a greater proportion of its momentum to the charged fluid through elastic interactions. Furthermore, the intensification of ionization processes during reconnection, driven by the localized thermal enhancement, facilitates an additional transfer of momentum and energy as neutral mass is converted into the charged fluid. 

Regarding the energy dissipated via Ohmic resistivity,  corresponding to the bottom-right panel of the figure \ref{fig:LoE}, we record values of $8.84 \times 10^{22}$, $1.07 \times 10^{23}$, and $1.27 \times 10^{23}$~erg for the $100$, $110$, and $120$~G cases, respectively. Evaluating the growth in dissipated energy relative to the $100$~G baseline reveals increases of $20.92\%$ and $43.89\%$ for the $110$ and $120$~G configurations. Notably, these variations align closely with the percentage increase in the initial magnetic energy stored within the system. This correlation suggests that Ohmic dissipation scales linearly with the available magnetic energy density ($e_M \propto B^2$), implying that the total resistive heating is directly proportional to the square of the initial magnetic field strength. 

Furthermore, while the declining reconnection rates might suggest the cessation of the primary reconnection event, the evolution of the magnetic energy and the cumulative Ohmic dissipation tends to plateau. This diminishing slope in their respective temporal variations indicates that the bulk conversion of magnetic energy is reaching completion. A similar trend is observed in the enthalpy of both species, which asymptotically approaches a stable state.

In contrast, the kinetic energy doesn't exhibit immediate stabilization; rather, it continues to increase even after the thermal and magnetic energy reservoirs have equilibrated. This persistent evolution is likely a signature of the chromospheric environment, characterized by higher plasma densities and higher plasma-$\beta$ regimes relative to the corona \cite{priest2002magnetic}. In such a dense, partially ionized medium, energy transformation is tempered by significant inertia and high collisionality. The hydrodynamic non-equilibrium of the region, coupled with the momentum stored in the neutral background, prolongs the acceleration phase. Consequently, the kinetic energy scales would likely continue to rise beyond the current $15$-second window, reaching even larger magnitudes if the simulation duration were extended.

Finally, the observed energy budget, comprising the liberated magnetic energy, the growth in kinetic energy and enthalpy, and the Ohmic dissipation, consistently exhibits scales between $10^{22}$ and $10^{23}$~erg. While these magnitudes are below the $10^{27}$--$10^{32}$~erg range typically associated with standard solar flares \cite{shibata2011solar}, this discrepancy does not preclude the possibility of achieving flare-scale energetics. This result must be interpreted within the context of our simulation's spatial constraints. As established in Equation \ref{eq:Energies}, the total energy content is inherently extensive, scaling linearly with the integration volume ($E \propto V$). 

Given that the characteristic spatial scales of solar flares range from $10^1$ to $10^3$~Mm \cite{shibata2011solar}, scaling our current domain to match these observed dimensions would yield a volumetric increase of several orders of magnitude. For instance, an increase of $10^1$ in the micro-flare characteristic length scale corresponds to a $10^3$ factor in volume. Consequently, at realistic micro-flare scales, the released energy would reach magnitudes of approximately $10^{26}$~erg. This positioning aligns our results with the energetic regime of micro-flares \cite{shibata2011solar}, suggesting that the underlying reconnection mechanism identified in this work is a viable candidate for driving small-scale solar eruptive events. 

\section{Discussions and Conclusions}
\label{sec:DaC}

In this study, we employed a MAGNUS-based extension of the two-fluid equations to simulate magnetic reconnection within chromospheric conditions and evaluate its consistency with solar flare energetics. The MAGNUS framework was adapted to resolve the coupled evolution of neutral and charged fluid species, incorporating both solar atmospheric stratification and inter-species collisional dynamics. Crucially, the implementation of a dedicated module for collisional energy exchange ensured the numerical stability of the system, allowing for a robust representation of the reconnection environment.

Morphologically, despite lower intensities than typical coronal events, the system manifests a magnetic arcade and plasma ejections, confirming consistency with flare-like physics. Thermal evolution corroborates this interpretation: heating bifurcates into an upper locus, associated with lobe formation, and a lower locus, where plasma accelerates toward denser atmospheric layers. The arcade topology is clearly evidenced by both current density and temperature profiles. Notably, reconnection drives significant localized heating, increasing peak regional temperatures by $18.49\%$–$58.46\%$, suggesting that small-scale reconnection events contribute substantially to the chromospheric and coronal energy budget.

Initially, the chromospheric plasma is recombination-dominated, evidenced by its respective orders of magnitude. During reconnection, both frequencies rise significantly; however, the ionization rate increases by two orders of magnitude relative to recombination, eventually peaking with and exceeding the intensity maxima. This dynamically enhances the activity of the charged species. These results align with observed X-ray emission traces at the arcade base, cusp, and the resulting lobe, consistent with established flare diagnostics \cite{vievering2021foxsi,fletcher2024solar}.

Analysis of the collision terms reveals negligible relative velocities between fluids compared to the system's global dynamics, confirming strong collisional coupling. Furthermore, thermal transfer diagnostics demonstrate that this fluid interaction is a primary driver of the observed heating. The acceleration and heating of the neutral species are essentially dependent on their interaction with the charged population. Because the neutral component exhibits more efficient heating due to its purely hydrodynamic nature, a significant temperature gradient ensues, facilitating a net thermal energy transfer from the neutral to the charged species.

Temporal analysis yields peak reconnection rates of $0.226$, $0.253$, and $0.279$ for the $100$, $110$, and $120$~G cases, respectively, attained at $4.23$, $4.03$, and $3.88$~s. This demonstrates that increasing the magnetic field strength enhances the peak reconnection rate and accelerates the onset of the phenomenon. Although these rates exceed the typical literature range ($0.01$–$0.1$), they remain physically plausible, residing within the same order of magnitude ($\sim 10^{-1}$). Furthermore, scaling the simulation domain to solar flare dimensions would likely moderate these rates and extend the characteristic timescale. Finally, theoretical frameworks support reconnection rates ($M_A$) exceeding $0.1$ \cite{pontin2022magnetic}, and collisional effects are known to facilitate accelerated reconnection dynamics \cite{jara2019kinetic}.

The released energy spans $10^{22}$–$10^{23}$~erg, scaling with the initial magnetic energy density ($E_B \propto B^2$). Although these magnitudes fall below those of standard solar flares, this discrepancy is an artifact of our spatially constrained simulation domain, approximately $1/5$ the scale of a micro-flare in $x$ and $z$, and $1/1000$ in $y$. Given the scale-invariance of the solver, extrapolating these results to observational dimensions yields energy releases of $10^{26}$–$10^{28}$~erg, aligning with empirical observations of solar events. Furthermore, kinetic energy accumulation persists beyond the reconnection peak. This delayed energy conversion is attributable to the chromospheric environment: the elevated mass density and high plasma-$\beta$ regime introduce significant inertia and kinetic pressure, which retard magnetic advection and modulate the rate of energy propagation.

Finally, our results demonstrate that increasing magnetic field strength correlates with higher peak temperatures, liberated energy, Ohmic resistivity, and ionization/recombination frequencies. However, this growth is linear, outpacing the proportional increase in initial magnetic energy. This indicates an enhanced conversion efficiency: the fraction of released energy rose from $14.92\%$ to $16.40\%$ between the $100$~G and $120$~G cases. Correspondingly, enthalpy and kinetic energy exhibited significant gains, with peak kinetic energy increasing by up to $129.82\%$. We attribute this heightened efficiency to collisional coupling, which facilitates rapid energy transfer from the neutral to the charged population. In the case of enthalpy, the neutral fluid acts as an energy reservoir, transferring thermal energy to the charged species, evidenced by a $15.11\%$ decrease in neutral enthalpy. This mechanism is similarly operative in the kinetic energy budget: the kinetic energy of the charged species surged from $53.21\%$ to $129.82\%$, significantly outperforming the $29.85\%$–$65.64\%$ increase observed in the neutral population.

It is important to acknowledge several limitations inherent to the present numerical study. The simulations were performed in a 2.5D configuration, employing a reduced computational domain focused primarily on the solar chromosphere. This imposes inherent limitations on the representation of fully three-dimensional turbulent dynamics and magnetic-field-line stochasticity. In addition, although the two-fluid equations for charged and neutral species, including both elastic and inelastic collisions, were solved to capture the essential physics of chromospheric reconnection, certain significant physical processes were necessarily omitted from the model. Specifically, the current setup does not account for radiative transfer or non-local thermal conduction. Consequently, while the simulation successfully reproduces the key morphological and energetic signatures of the reconnection process, the resulting temperature evolution and energy budget remain subject to the limitations imposed. Future investigations will extend this work to full 3D simulations, incorporating a more comprehensive treatment of radiative and thermal effects.

In summary, this work demonstrates that chromospheric magnetic reconnection is a highly efficient process, driven by strong collisional coupling between neutral and charged fluid populations. Our MAGNUS-based simulations reveal that energy release is not merely a product of magnetic field decay, but is dynamically enhanced by efficient thermal energy transfer and kinetic acceleration. By bridging the gap between small-scale simulation energetics and observable micro-flare signatures, these results provide a theoretical framework for understanding the role of partially ionized plasmas in the heating of the solar atmosphere. These findings underscore the importance of accounting for multi-species interactions in future models of solar eruptive events.

\ack{We thank the anonymous referee for their valuable comments and suggestions. F.D.L-C was supported by the Vicerrectoría de Investigación y Extensión - Universidad Industrial de Santander, under Grant No. 3703.}



\section*{References}

\providecommand{\newblock}{}

\end{document}